\documentclass[12pt,]{article}
 \usepackage{graphicx}
 \usepackage[cp1251]{inputenc}
 \usepackage{epstopdf}
 \usepackage{rotating}
\begin{document}

 \bigskip

 \bigskip

 \centerline {\bf A NEW CATALOG OF ORBITS OF 152 GLOBULAR }
 \centerline {\bf CLUSTERS FROM Gaia EDR3}

  \bigskip

  \bigskip

   \centerline{\bf A. T. Bajkova, V. V. Bobylev }

  \bigskip
  \bigskip
  \centerline{\small \it Main (Pulkovo) Astronomical Observatory of the Russian Academy of Sciences}
  \bigskip

  \bigskip

{{\bf Abstract.}
This paper provides a new catalog of orbits and their parameters for a practically complete list of currently known galactic globular clusters (GCs), compiled by Vasiliev (2019) based on the most accurate modern measurements of their velocities and positions. The integration of the orbits of 152 globular clusters for 5 Gyr backward was performed using the new average proper motions obtained from the Gaia EDR3 catalog (Vasiliev and Baumgardt, 2021) and new average distances (Baumgardt and Vasiliev, 2021) in the axisymmetric three-component potential with spherical bulge, disk component, and spherical dark Navarro-Frank-White halo (Bajkova and Bobylev, 2016). The new orbital parameters are compared with the orbital parameters constructed by us earlier (Bajkova and Bobylev, 2021) in the same gravitational potential using proper motions obtained from the Gaia DR2 catalog (Vasiliev, 2019) and with the distances from the Harris catalog (2010).
}

  \bigskip


\newpage
\section*{Introduction}
Globular clusters (GCs) are among the most interesting objects in our Galaxy. Their study allows us to understand the birth and evolution of the Galaxy, since they are the oldest stellar formations. Their age is almost equal to the age of the Universe. Currently, about 170 GCs of the Milky Way are known. According to theoretical estimates, the number of GCs in the Milky Way can be on the order of 200.

One of the methods for investigation of GCs is to study their orbital motion, which has become possible
thanks to high-precision measurements of their spatial velocities and positions from the Gaia spacecraft. The appearance of catalogs of mean proper motions already according to the data of the second release of DR2 in combination with other astrometric data on radial velocities and positions of GCs made it possible to study the orbital motion of almost all currently known GCs (for example, Helmi et al., 2018; Baumgardt et al., 2019; Vasiliev, 2019; Bajkova et al., 2020; Bajkova and Bobylev, 2021).

Among the catalogs of astrometric data with proper motions from Gaia DR2 (Brown et al., 2018), we especially note the catalog by Vasiliev (2019) for about 150 GCs, which allows one to build the 6D phase space needed to calculate the orbits. This catalog was used by us to study the orbital properties of GCs, and on this basis, we developed a new method for dividing GCs into subsystems of the Galaxy: bulge, thick disk, and halo (Bajkova et al., 2020), based on the bimodal nature of the distribution of the $L_Z/ecc$ parameter , where $L_Z$ is the $Z$   component of the angular momentum, $ecc$ is the eccentricity of the orbit.

In the work of Bajkova and Bobylev (2021), we present a catalog of the orbits of 152 GCs and their orbital parameters, and also propose a modified classification by Massari et al. (2019) according to subsystems of the Galaxy, based on the obtained orbital properties of GCs.

With the advent of a new, more accurate version of the GCs proper motion catalog (Vasiliev and Baumgardt, 2021) based on Gaia EDR3 measurement data (Brown et al., 2021), as well as new average GCs distances (Baumgardt and Vasiliev, 2021), a natural problem arises of constructing a new catalog of GCs orbits and refining their parameters. This work is dedicated to this aim.

The work is structured as follows. In the first section, a brief description and substantiation of the adopted model of the gravitational potential is given, in which the GCs orbits are integrated, the equations of motion are given, and formulas are given for calculating the orbital parameters. The second section describes the data, compares the average proper motions and their uncertainties obtained from the data of the Gaia DR2 and EDR3 catalogs, and compares the new GCs distances with the previously used distances from the Harris (2010) catalog. The third section is devoted to a presentation of the results of the work. A catalog of GCs orbits and their parameters calculated from new data is given, a comparison of the main orbital parameters with the parameters published in the work of Bajkova and Bobylev (2021) is made. The Conclusion gives a short summary of the  main results of the work.

\section{Method}
\subsection{Model of axisymmetric galactic potential}
The axisymmetric gravitational potential of the Galaxy is presented as the sum of three components: the central spherical bulge $\Phi_b(r(R,Z))$, the disk $\Phi_d(r(R,Z))$, and the massive spherical dark matter halo $\Phi_h( r(R,Z))$ (see Bajkova and Bobylev (2016) and references therein):
\begin{equation}
\begin{array}{lll}
  \Phi(R,Z)=\Phi_b(r(R,Z))+\Phi_d(r(R,Z))+\Phi_h(r(R,Z)).
 \label{pot}
 \end{array}
 \end{equation}
Here we use a cylindrical coordinate system ($R,\psi,Z$) with origin at the center of the Galaxy. In a rectangular Cartesian coordinate system $(X,Y,Z)$ with origin at the center of the Galaxy, the distance to the star (spherical radius) is $r^2=X^2+Y^2+Z^2=R^2+Z^ 2$.
The gravitational potential is expressed in units of 100 km$^2$ s$^{-2}$, distances - in kpc, masses - in units of the mass of the Galaxy, $M_{0}=2.325\times 10^7 M_\odot$ , the gravitational constant is $G=1.$

The potentials of the bulge $\Phi_b(r(R,Z))$ and the disk $\Phi_d(r(R,Z))$ are expressed in the form proposed by Miyamoto and Nagai (1975):
 \begin{equation}
  \Phi_b(r)=-\frac{M_b}{(r^2+b_b^2)^{1/2}},
  \label{bulge}
 \end{equation}
 \begin{equation}
 \Phi_d(R,Z)=-\frac{M_d}{\Biggl[R^2+\Bigl(a_d+\sqrt{Z^2+b_d^2}\Bigr)^2\Biggr]^{1/2}},
 \label{disk}
\end{equation}
where $M_b, M_d$ are the masses of these components, and $b_b, a_d, b_d$ are the scale lengths of the components in kpc.

To describe the halo component, we used the Navarro-Frank-White (NFW) expression presented in Navarro et al. (1997):
 \begin{equation}
  \Phi_h(r)=-\frac{M_h}{r} \ln {\Biggl(1+\frac{r}{a_h}\Biggr)},
 \label{halo-III}
 \end{equation}
where $M_h$ is the mass, $a_h$ is the scale length.

\begin{figure*}
\begin{center}
   \includegraphics[width=0.5\textwidth,angle=-90]{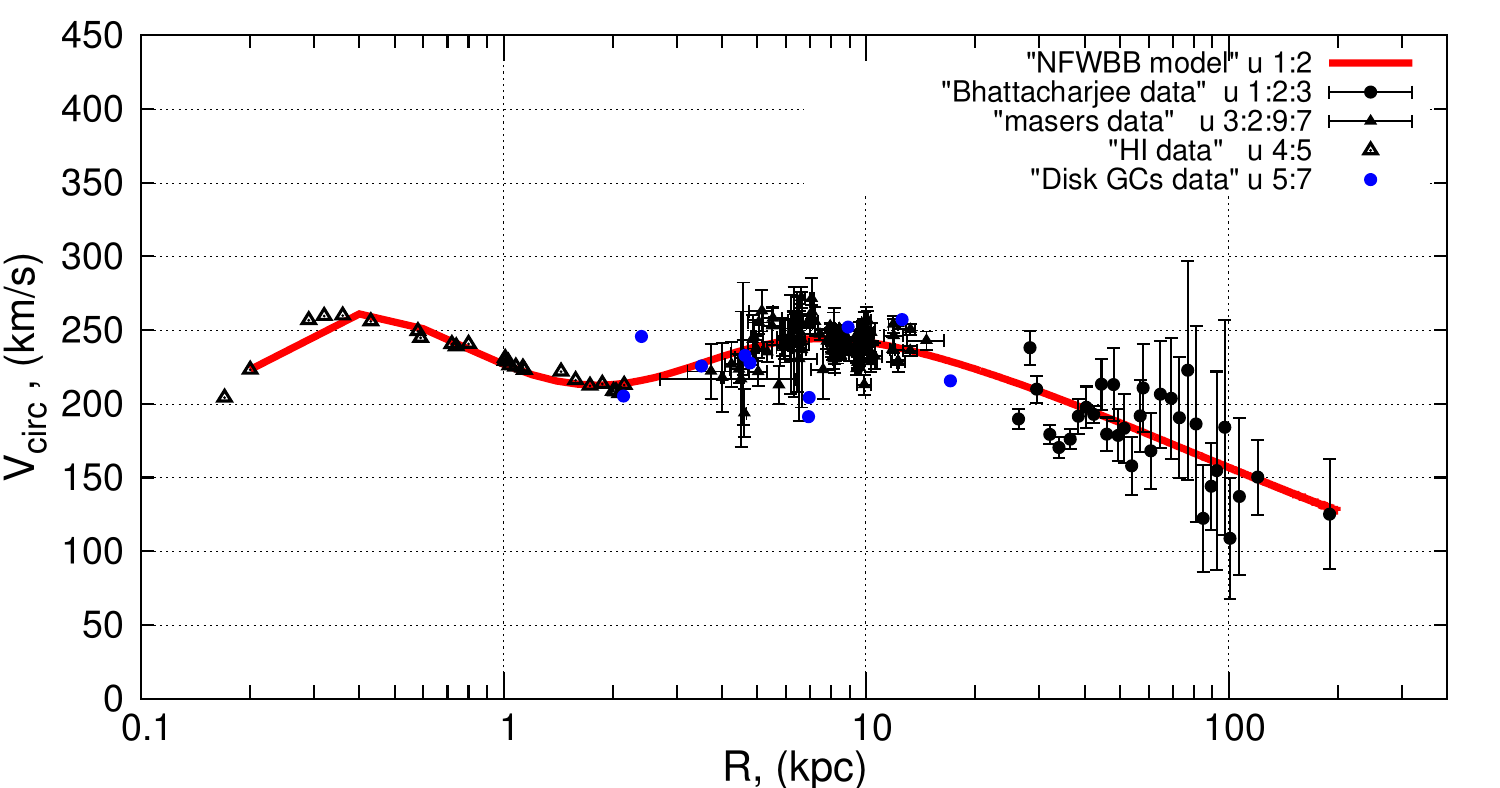}
  \caption{Rotation curve corresponding to the NFWBB potential model. The blue dots show the circular velocities of the disk GCs with orbital eccentricities $<0.2$.}
\label{rotc}
\end{center}
\end{figure*}

The model of the galactic potential adopted by us, which for brevity we denote as NFWBB, has parameters obtained as a result of their fitting to the data on the circular velocities of clouds of ionized hydrogen HI, maser sources, and various halo objects with large galactocentric distances $R$ up to $\sim200$ kpc from Bhattacharjee et al. (2014) (see Fig.~\ref{rotc}). In addition, when adjusting the parameters, restrictions were used on the local dynamic density of matter $\rho_\odot=0.1 M_\odot$~pc$^{-3}$ and the force acting perpendicular to the plane of the Galaxy $|K_{z=1.1}| /2\pi G=77M_\odot$~pc$^{-2}$ (Irrgang et al., 2013).

NFWBB model parameters are given in Table~\ref{t:model}. The corresponding model rotation curve up to distances $R=200$ kpc is shown in Fig.~\ref{rotc}. When constructing the rotation curve, we used the values $R_\odot=8.3$ kpc for the galactocentric distance of the Sun and $V_\odot=244$ km s$^{-1}$ for the linear velocity of rotation of the Local Standard of Rest around the center of the Galaxy, as adopted by Bhattacharjee et al ( 2014). The mass of the Galaxy according to this model (Bajkova and Bobylev 2016) is equal to $M_{G_{(R \leq 200~kpc)}}=0.75\pm0.19\times10^{12}M_\odot$. This value is in good agreement with some modern independent estimates. So, for example, the lower estimate of the mass of the NFW halo, obtained quite recently by Koppelman, Helmi (2021) from the data on the velocities of runaway halo stars, is $M_{G_{(R \leq 200~kpc)}}=0.67^{+0.30} _{-0.15}\times10^{12}M_\odot$.
In Fig.~\ref{rotc}, in addition to the available data, we also plot (blue dots) the circular velocities of the thick disk GCs identified by us with orbital eccentricities $<0.2$, which show good agreement with data on maser sources in the interval of galactocentric distances $ 2 < R < $20 kpc.

The NFWBB model of the gravitational potential of the Milky Way seems to us the most realistic compared to other known models, since it is supported by data at large galactocentric distances, which is very important when integrating the orbits of distant globular clusters and clusters with large apocentric distances, and also gives good agreement with modern estimates local parameters and a number of independent estimates of the mass of the Galaxy (Bajkova and Bobylev, 2017), a thorough review of which is also given in the recent paper by Wang et al. (2020).

 \begin{table}
 \begin{center}
 \caption[]
 {\baselineskip=1.0ex
Potential model parameters, $M_0=2.325\times10^7 M_\odot$}
  \bigskip
  \bigskip
 \label{t:model}
 \begin{tabular}{|l||r|}\hline
 Parameter     &     Value         \\\hline
 $M_b$ [$M_0$] &    443$\pm27$     \\
 $M_d$ [$M_0$] &   2798$\pm84$     \\
 $M_h$ [$M_0$] &  12474$\pm3289$   \\
 $b_b$ [kpc]   & 0.2672$\pm0.0090$ \\
 $a_d$ [kpc]   &   4.40$\pm0.73$   \\
 $b_d$ [kpc]   & 0.3084$\pm0.0050$ \\
 $a_h$ [kpc]   &    7.7$\pm2.1$    \\\hline
 $R_\odot$ [kpc]    &  8.30 \\\hline
 $V_\odot$ [km s$^{-1}$] & 243.9 \\\hline
 $M_{G_{(R \leq 200 kpc)}}$ & 0.75$\pm$0.19  \\
  $[10^{12} M_\odot]$     &               \\\hline
 \end{tabular}
  \end{center}
  \end{table}

\subsection{Orbit integration}
The equation of motion of a test particle in an axisymmetric gravitational potential can be obtained from the Lagrangian of the $\pounds$ system (see Appendix A in Irrgang et al. (2013)):
\begin{equation}
 \begin{array}{lll}
 \pounds(R,Z,\dot{R},\dot{\psi},\dot{Z})=\\
 \qquad0.5(\dot{R}^2+(R\dot{\psi})^2+\dot{Z}^2)-\Phi(R,Z).
 \label{Lagr}
 \end{array}
\end{equation}
Introducing the canonical moments
\begin{equation}
 \begin{array}{lll}
    p_{R}=\partial\pounds/\partial\dot{R}=\dot{R},\\
 p_{\psi}=\partial\pounds/\partial\dot{\phi}=R^2\dot{\psi},\\
    p_{Z}=\partial\pounds/\partial\dot{Z}=\dot{Z},
 \label{moments}
 \end{array}
\end{equation}
we obtain the Lagrange equations in the form of a system of six first-order differential equations:
 \begin{equation}
 \begin{array}{llllll}
 \dot{R}=p_R,\\
 \dot{\psi}=p_{\psi}/R^2,\\
 \dot{Z}=p_Z,\\
 \dot{p_R}=-\partial\Phi(R,Z)/\partial R +p_{\psi}^2/R^3,\\
 \dot{p_{\psi}}=0,\\
 \dot{p_Z}=-\partial\Phi(R,Z)/\partial Z.
 \label{eq-motion}
 \end{array}
\end{equation}

To integrate the equations (\ref{eq-motion}), we used the fourth-order Runge-Kutta algorithm. Integration was carried out 5 Gyr backward. As shown by Bajkova et al. (2021), the galactic potential can be considered stationary in this time interval.

The peculiar velocity of the Sun relative to the Local Standard of Rest was taken equal to
$(u_\odot,v_\odot,w_\odot)=(11.1,12.2,7.3)\pm(0.7,0.5,0.4)$ km s$^{-1}$ (Sch\"onrich et al., 2010). Here we use heliocentric velocities in a moving Cartesian coordinate system with the velocity $u$ directed towards the Galactic center, $v$ in the direction of the Galaxy's rotation and $w$ perpendicular to the plane of the Galaxy and directed towards the north pole of the Galaxy.

Let the initial positions and spatial velocities of the test particle in the heliocentric coordinate system be equal to $(x_o,y_o,z_o,u_o,v_o,w_o)$. Then the initial positions ($X,Y,Z$) and velocities ($U,V,W$) of the test particle in Cartesian galactic coordinates are given by the formulas:
\begin{equation}
 \begin{array}{llllll}
 X=R_\odot-x_o, Y=y_o, Z=z_o+h_\odot,\\
 R=\sqrt{X^2+Y^2},\\
 U=u_o+u_\odot,\\
 V=v_o+v_\odot+V_0,\\
 W=w_o+w_\odot,
 \label{init}
 \end{array}
\end{equation}
where $R_\odot$ and $V_\odot$ are the Galactocentric distance and linear velocity of rotation of the Local Standard of Rest around the center of the Galaxy, $h_\odot=16$~pc (Bobylev and Bajkova, 2016) is the height of the Sun above the Galactic plane, $\Pi$ and $\Theta$ are the radial and circular velocities, respectively.

In this work, we calculate the following parameters of the orbits of globular clusters using well-known formulas (the units of measurement of the parameters are given in Table~2):

\medskip

\noindent (1)initial distance of the GC from the center of the Galaxy $d_{GC}$:
\begin{equation}
d_{GC}=\sqrt{X^2+Y^2+Z^2};
\end{equation}

\medskip

\noindent (2)radial velocity $\Pi$:
\begin{equation}
 \Pi=-U \frac{X}{R}+V \frac{Y}{R};
 \end{equation}

\medskip

\noindent (3) circular velocity $\Theta$:
\begin{equation}
 \Theta=U \frac{Y}{R}+V \frac{X}{R};
 \end{equation}

\medskip

\noindent (4) total 3D velocity $V_{tot}$:
\begin{equation}
V_{tot}=\sqrt{\Pi^2+\Theta^2+W^2};
 \end{equation}

\medskip

\noindent (5)apocentric distance (apo) of the orbit;

\medskip

\noindent (6)pericentric distance (peri) of the orbit;

\medskip

\noindent (7)eccentricity (ecc) of the orbit:
\begin{equation}
ecc=\frac{apo-peri}{apo+peri};
 \end{equation}

\medskip

\noindent (8)angular momentum components:
\begin{equation}
 L_X=Y\times W-Z\times V;
 \end{equation}

 \begin{equation}
 L_Y=Z\times U-X\times W;
 \end{equation}

\begin{equation}
 L_Z=X\times V-Y\times U;
 \end{equation}

\medskip

\noindent (9) orbital inclination $\theta$:
\begin{equation}
\theta=\arccos(\frac{L_Z}{L}),
\end{equation}
where $L=\sqrt{L_X^2+L_Y^2+L_Z^2}$ is total orbital momentum;

\medskip

\noindent (10)orbit period $T_r$;

\medskip

\noindent (11)total energy $E$:
\begin{equation}
 E= \Phi(R,Z)+\frac{V_{tot}^2}{2}.
 \end{equation}

The uncertainties of the orbital parameters were calculated by the Monte Carlo method using 100 iterations, taking into account the uncertainties in the initial coordinates and GC velocities, as well as the errors in the peculiar velocity of the Sun.

\section{Data}
For the 152 globular clusters we studied earlier (Bajkova and Bobylev, 2021) with data mainly from the Vasiliev (2019) catalog, we took new average values of proper motions and their uncertainties from the new Vasiliev and Baumgardt (2021) catalog obtained from the Gaia EDR3 catalog data , as well as new average distances from Baumgardt and Vasiliev (2021). All other astrometric data (radial velocities, coordinates) remained the same.

\begin{figure*}
{\begin{center}
   \includegraphics[width=0.35\textwidth,angle=-90]{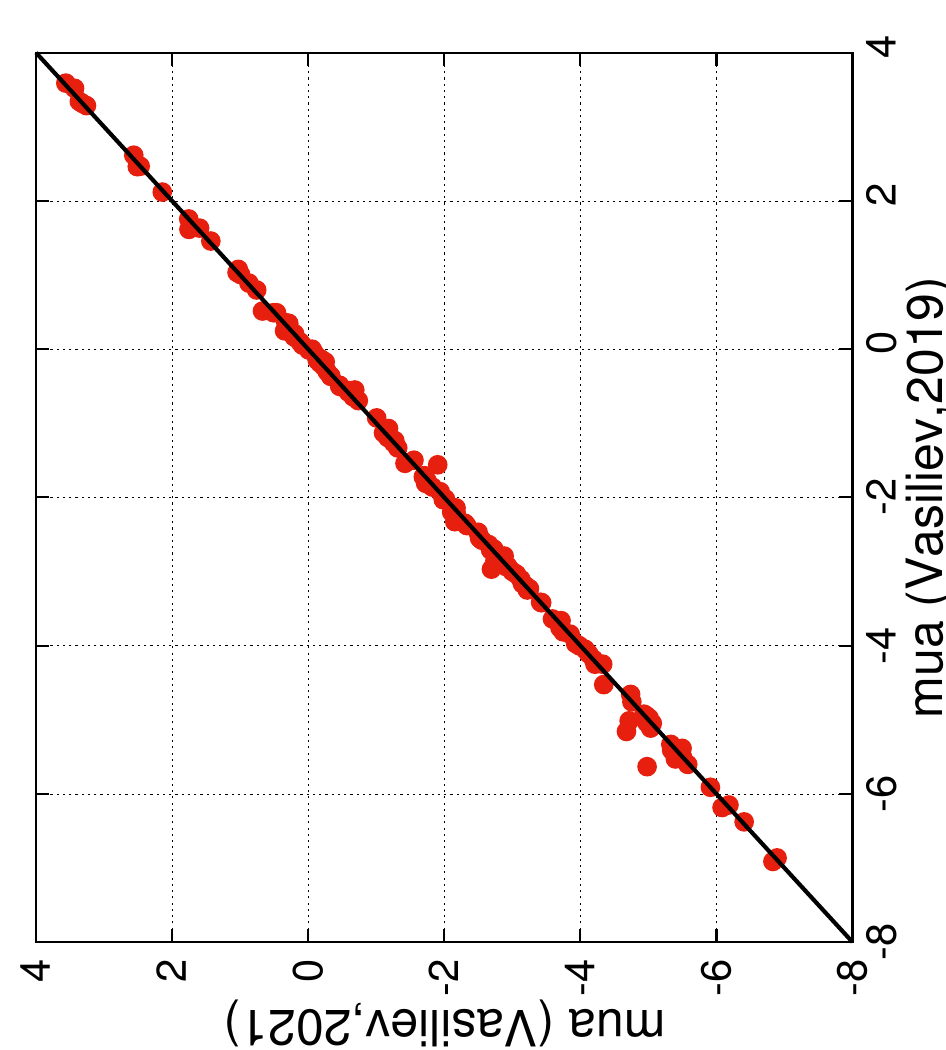}
   \includegraphics[width=0.35\textwidth,angle=-90]{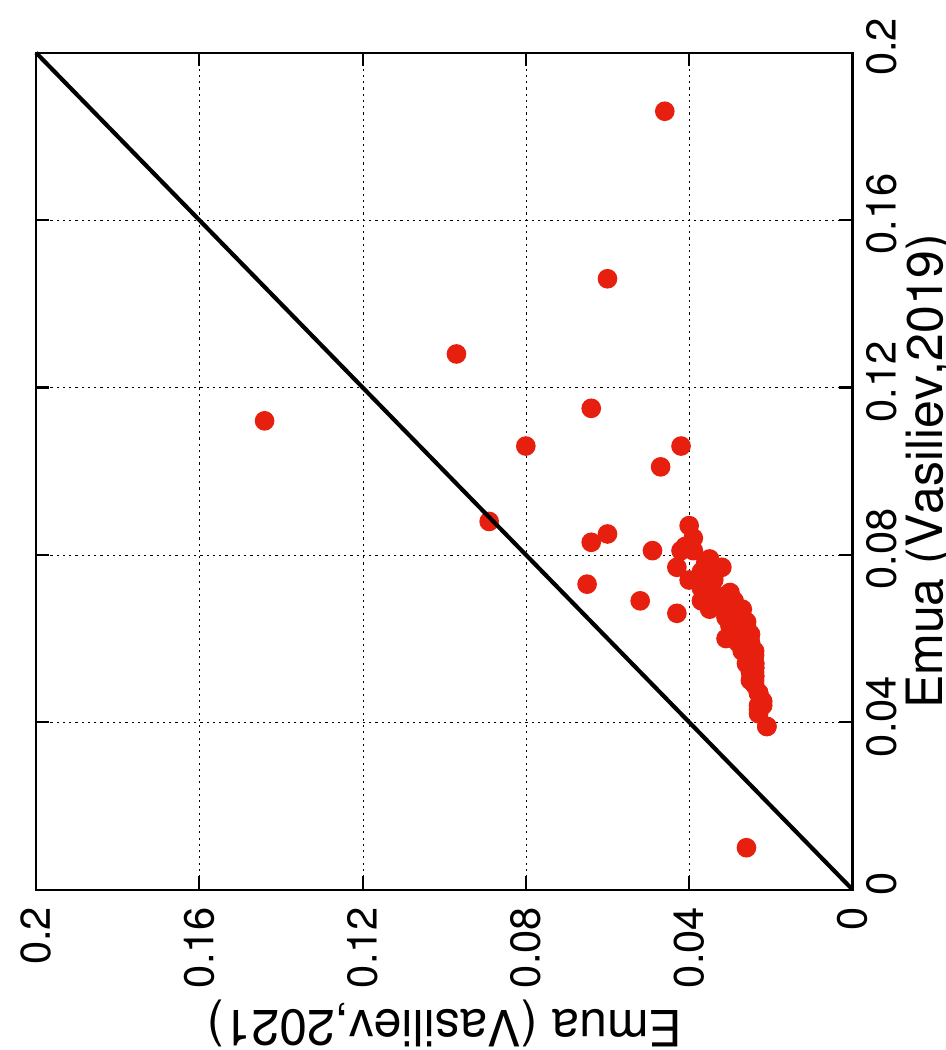}\\
   \includegraphics[width=0.35\textwidth,angle=-90]{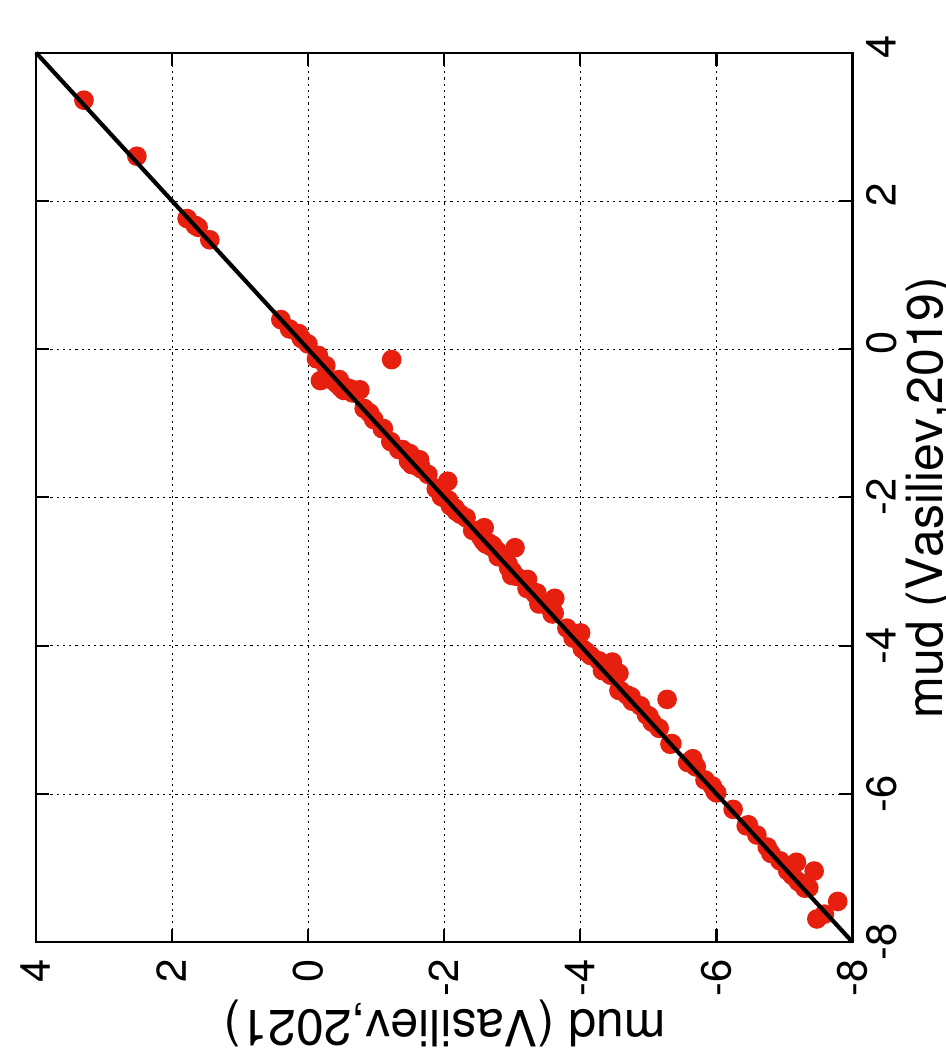}
   \includegraphics[width=0.35\textwidth,angle=-90]{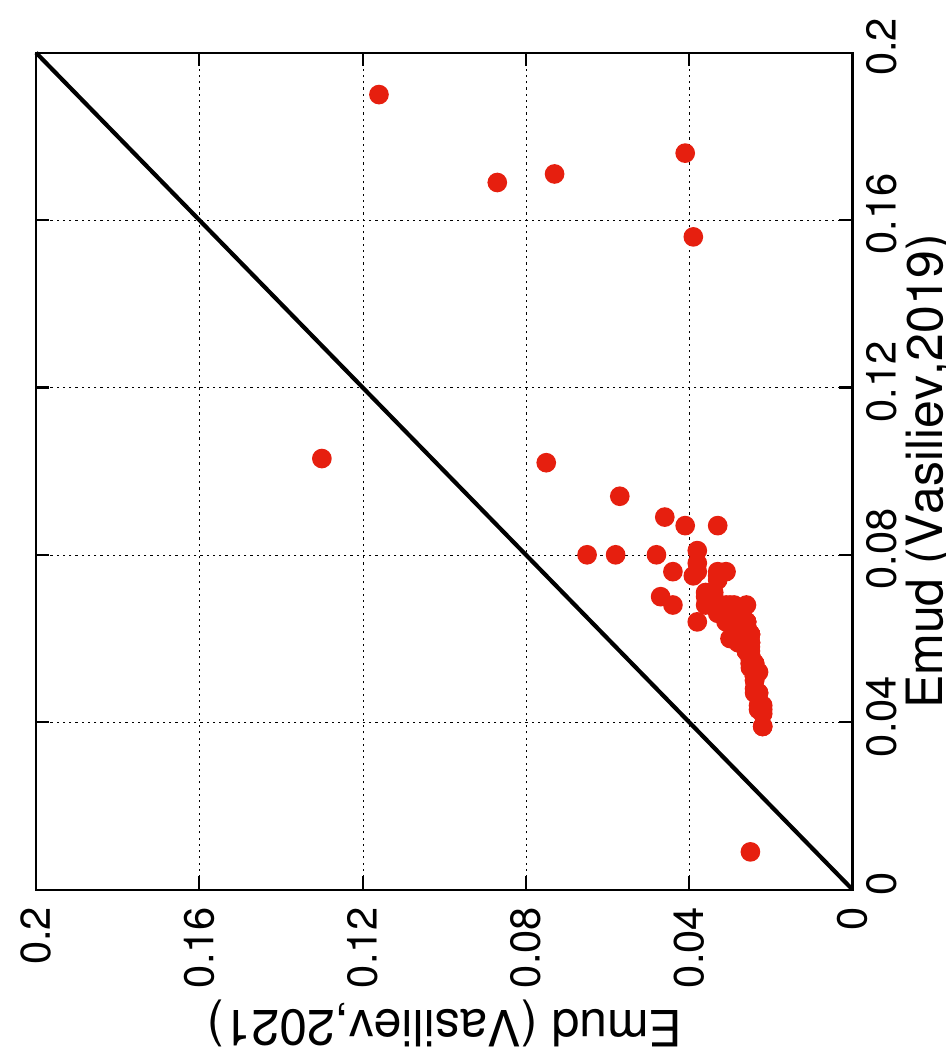}\\
  \caption{Comparison of GC proper motions (in $\alpha$ ($\mu_\alpha$) and $\delta (\mu_\delta$), indicated on the graphs as mua and mud, respectively) and their uncertainties (Emua) and (Emud) , respectively) from the Vasiliev (2019) catalog (Gaia DR2, horizontal axis)) and the Vasiliev and Baumgardt (2021) catalog (Gaia EDR3, vertical axis). Each panel has a matching line.}
\label{fmuad}
\end{center}}
\end{figure*}

In Fig.~\ref{fmuad} we give a comparison of the mean proper motions from these two catalogs obtained from measurements of Gaia DR2 and Gaia EDR3. As follows from the figure, the new values of proper motions for a number of GCs noticeably differ from the old ones. At the same time, the accuracy of measuring new proper motions increased on average by a factor of two.

\begin{figure*}
{\begin{center}
    \includegraphics[width=0.35\textwidth,angle=-90]{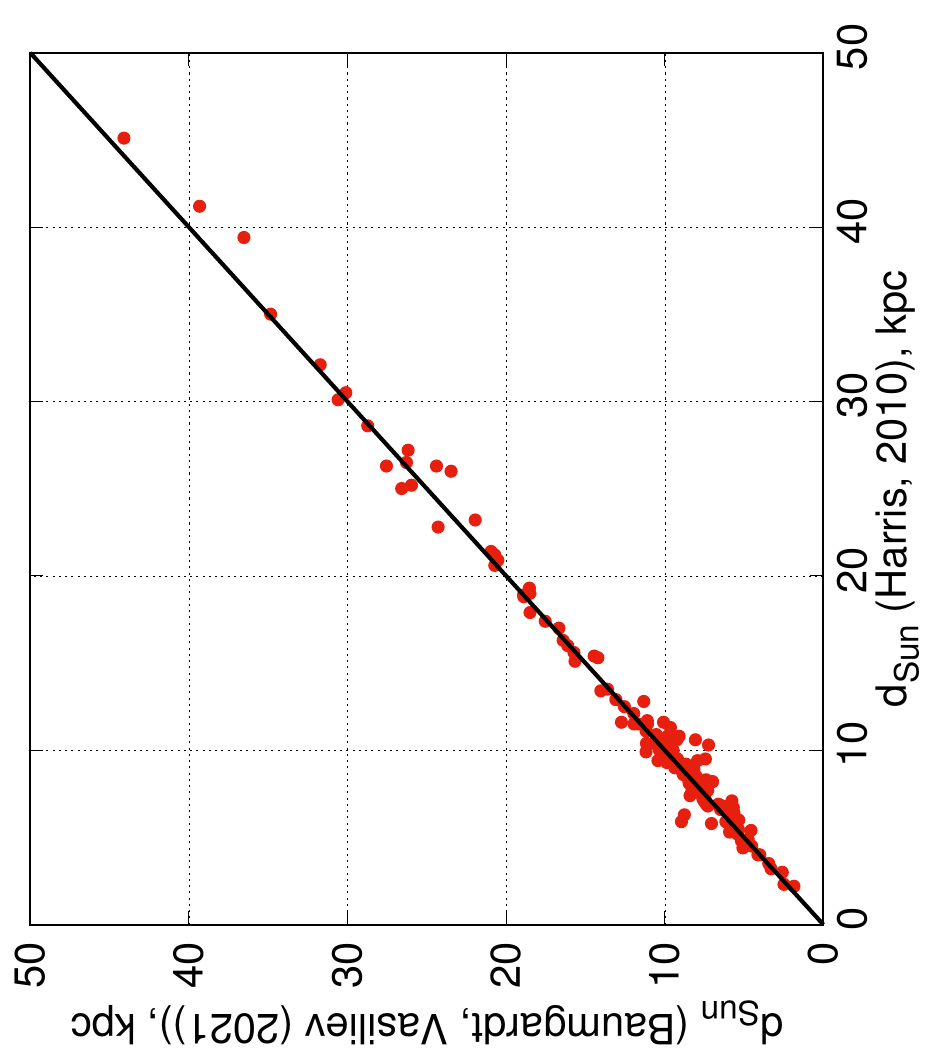}
   \includegraphics[width=0.35\textwidth,angle=-90]{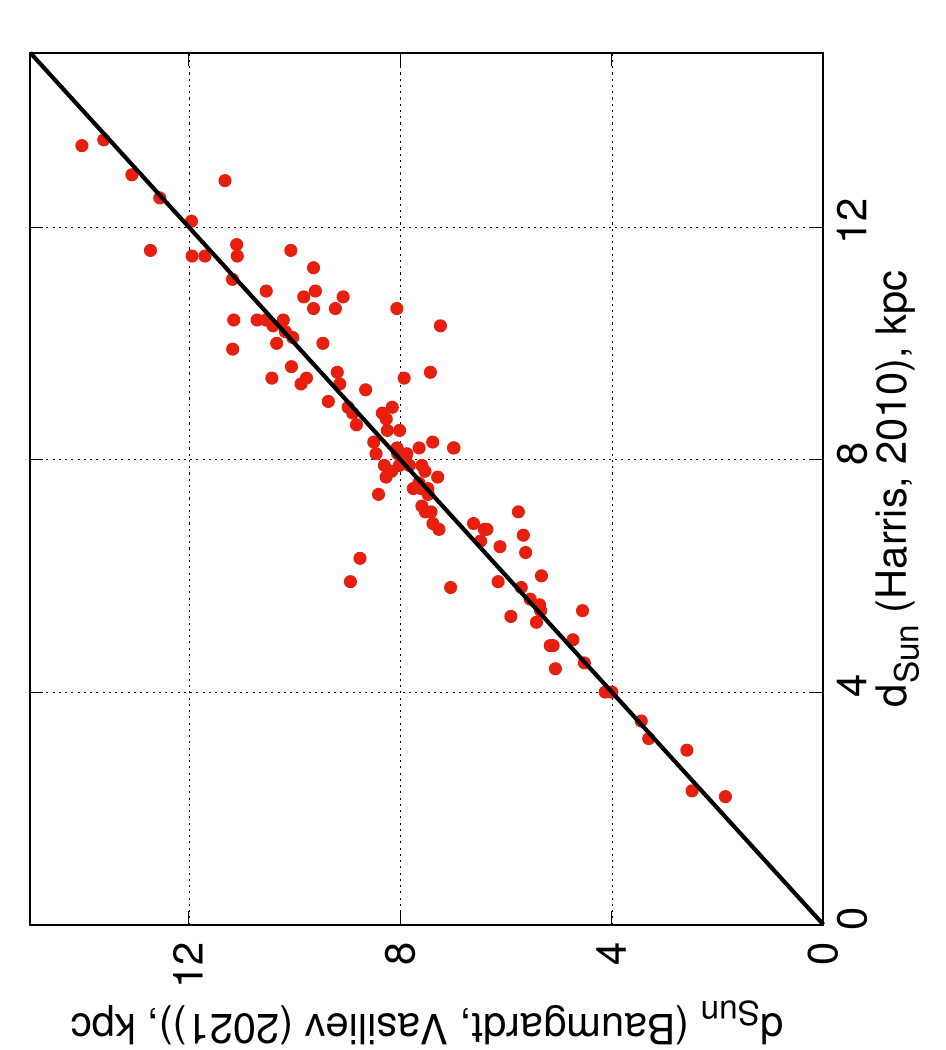}\
  \caption{Comparison of GC heliocentric distances ($d_{Sun}$) from the Harris (2010) catalog (horizontal axis) and the Baumgardt and Vasiliev (2021) catalog (vertical axis). The left panel shows distances up to 50 kpc, the right panel shows distances up to 15 kpc. Each panel has a matching line.}
\label{fdist}
\end{center}}
\end{figure*}

Figure~\ref{fdist} compares the GC distances to the Sun, taken from the Harris (2010) catalog when compiling the previous catalog of orbits, and the average distances from the Baumgardt and Vasiliev (2021) catalog, which we used when compiling the new catalog. The left panel reflects distances up to 50 kpc, while the right panel reflects distances up to 15 kpc. From the above figures (especially for small $d_{Sun}$), it can be seen that the distances for a number of GCs differ quite significantly, which, together with changes in proper motions, as will be shown below, significantly affected the orbital motion (and, accordingly, the orbital parameters) of many GCs.

\section{Results}
Table~2 gives the orbital parameters $d_{GC}$, $\Pi$, $\Theta$, $V_{tot}$, $apo$, $peri$, $ecc$, $\theta$, $T_r$, $L_Z$, $E$ (see Section 1.2) of 152 GCs calculated for the new EDR3 average proper motions (Vasiliev and Baumgardt 2021) as well as new average distances (Baumgardt and Vasiliev, 2021).

\begin{figure*}
{\begin{center}
   \includegraphics[width=1.0\textwidth,angle=0]{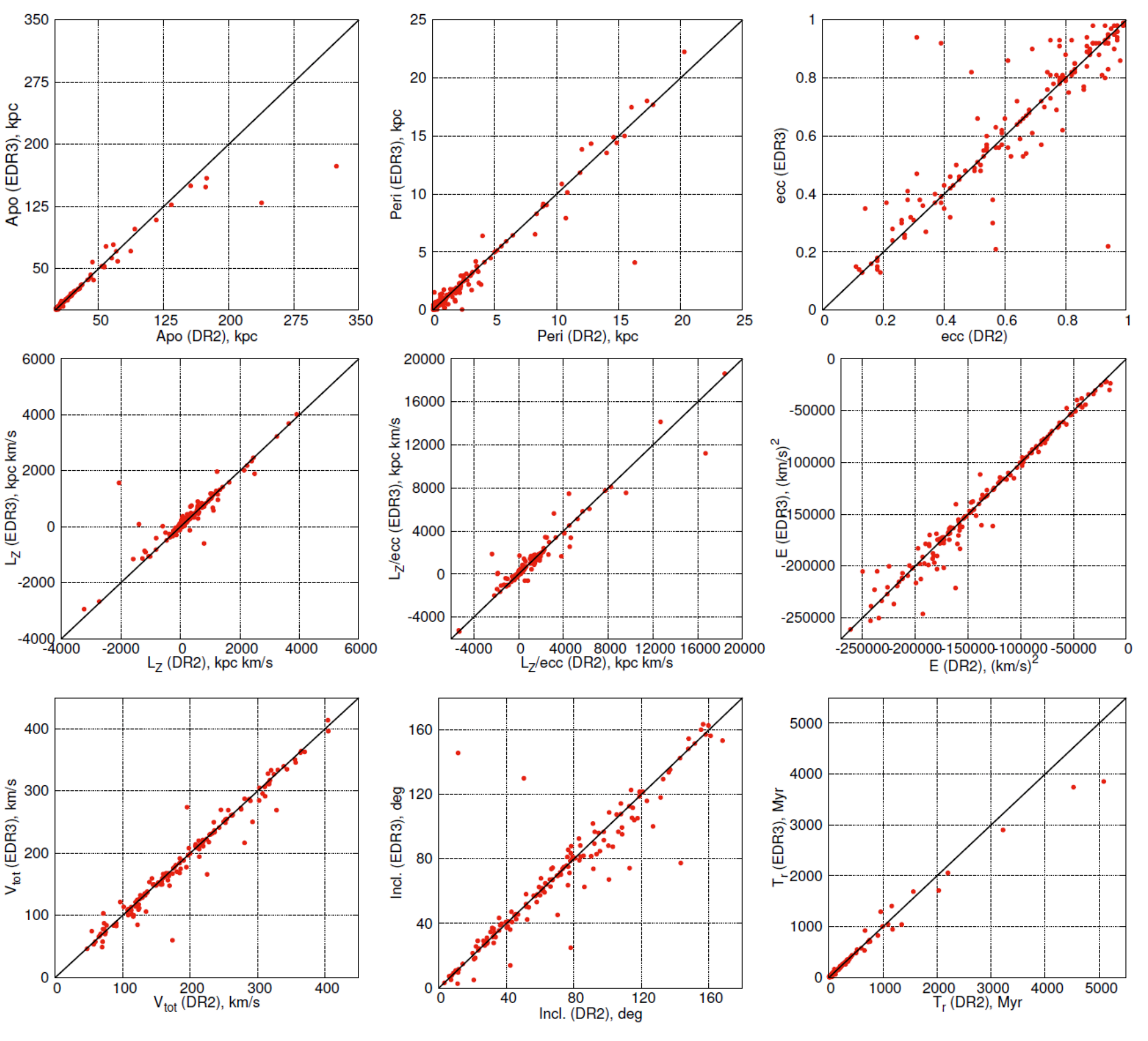}
    \caption{Comparison of GC orbit parameters ($apo, peri, ecc, L_Z, L_Z/ecc, E, V_{tot}, \theta, T_r$) obtained from the DR2 catalog (horizontal axis) and the EDR3 catalog (vertical axis). Each panel has a matching line.}
\label{fcomp}
\end{center}}
\end{figure*}

A comparison of the main orbital parameters of the GC calculated using the average proper motions of Gaia EDR3 with similar parameters calculated from the data of the Gaia DR2 catalog is shown in Fig.~\ref{fcomp}. As can be seen from the figures, a significant difference is observed for a number of GCs.

As the analysis shows, the most distant objects and objects with highly elongated orbits in the radial direction have undergone the greatest change in their orbital properties. We also note that the classification of Massari (2019) of GCs by subgroups, modified by Bajkova and Bobylev (2021), has been preserved, since the parameter $L_Z/ecc$, as can be seen from the figure, has not undergone major changes.

The new catalog of orbits in two projections $(X,Y)$ and $(R,Z)$ of 152 globular clusters is shown in Fig.~5, which can be compared with the previous one published by Bajkova and Bobylev (2021).

\section{Conclusion}
The emergence of more and more accurate astrometric data on the coordinates and spatial velocities of globular clusters makes it possible to study their motion in three-dimensional space by integrating orbits in the gravitational potential of the Galaxy.

Already thanks to the Gaia DR2 data (Helmi et al., 2018; Baumgardt et al., 2019; Vasiliev, 2019) on the proper motions of almost all currently known globular clusters, it became possible to study their kinematics and dynamics, and to classify GCs by subsystems of the Milky Way in order to determine objects that were formed directly in the Galaxy, or introduced from outside as a result of accretion from other (dwarf) galaxies surrounding the Milky Way (Massari et al., 2019). The creation of a catalog of orbits and their parameters for almost all GCs (Bajkova and Bobylev, 2021) with known data on the 6D phase space required for orbit integration provides highly informative material for subsequent research.

The recent appearance of a new, more accurate version of the catalog of proper motions (Vasiliev and Baumgardt, 2021), as well as new, more accurate average distances (Baumgardt and Vasiliev, 2021), made it possible to create a new version of the catalog of orbits of 152 GCs and their parameters, which was the result of this work.

The authors are grateful to the referee for useful remarks that contributed to the improvement of the article.

 \bigskip\medskip{REFERENCES}\medskip {\small
 \begin{enumerate}

\item
A.T. Bajkova and V.V. Bobylev, Astron. Lett. {\bf 42}, 567 (2016). 

\item

A.T. Bajkova and V.V. Bobylev, OAst. {\bf 26}, 72 (2017). 

\item
A.T. Bajkova, G. Carraro, V.I. Korchagin, N.O. Budanova, and V.V. Bobylev, Astrophys. J. {\bf 895}, 69 (2020).

\item
A.T. Bajkova and V.V. Bobylev, RAA, {\bf 21}, Issue 7, 173 (2021).

\item
A.T. Bajkova, A.A.Smirnov, and V.V. Bobylev, Astron. Lett. {\bf 47}, 454 (2021).

\item
H. Baumgardt, M. Hilker, A. Sollima, and A. Bellini, MNRAS {\bf 482}, 5138 (2019).

\item
H. Baumgardt and E. Vasiliev, MNRAS {\bf 505}, 5957 (2021).

\item
P. Bhattacharjee, S. Chaudhury, and S. Kundu, Astrophys. J. {\bf 785}, 63 (2014).

\item
V.V. Bobylev and A.T. Bajkova, Astron. Lett. {\bf 42}, 1 (2016).     

\item
Gaia Collaboration, A.G.A. Brown, A. Vallenari, T. Prusti, et al., Astron. Astrophys. {\bf 616}, 1 (2018). 

\item
Gaia Collaboration, A.G.A. Brown, A. Vallenari, T. Prusti, et al., Astron. Astrophys. {\bf 649}, 1 (2021). 

\item
W. Harris, astro-ph/1012.3224 (2010).

\item
Gaia Collaboration: A. Helmi, F. van Leeuwen, P.J. McMillan, et al., Astron. Astrophys. {\bf 616}, 12 (2018).

\item
A. Irrgang, B. Wilcox, E. Tucker, and L. Schiefelbein, Astron. Astrophys. {\bf 549}, 137 (2013).

\item
H.H. Koppelman and A. Helmi, Astron. Astrophys. {\bf 649}, Id. A136, 14 pp. (2021).

\item
D. Massari, H.H. Koppelman, and A. Helmi, Astron. Astrophys. {\bf 630}, L4 (2019).

\item
M. Miyamoto and R. Nagai, PASJ {\bf 27}, 533 (1975).

\item
J.F. Navarro, C.S. Frenk, and S.D.M. White, Astrophys. J. {\bf 490}, 493 (1997).

\item
R. Sch\"onrich, J. Binney, and W. Dehnen, MNRAS  {\bf 403}, 1829 (2010).

\item
E. Vasiliev, MNRAS {\bf 484}, 2832 (2019).

\item
E. Vasiliev and H. Baumgardt, MNRAS {\bf 505}, 5978 (2021).

\item
W. Wang, J. Han, M. Cautun, Z. Li, and M. Ishigaki, SCPMA, {\bf 63}, id.109801 (2020).

\end{enumerate}
}

\newpage
\rotatebox{90}{
		\begin{minipage}{1.5\linewidth}
		\label{t:prop}
		\begin{scriptsize} \begin{center}
{\scriptsize {Table 2:} Orbital properties of the GCs. For each GC
the parameter values are obtained as a result of integration of the orbit for 5 Gyr backward.}
\medskip
\begin{tabular}{|l|r|r|r|r|r|r|r|r|r|r|r|}\hline
 Name &$d_{GC}$&$\Pi$ &$\Theta$ &$V_{tot}$ &apo   & peri      &ecc&incl. $\theta$    &$T_r$&$L_Z$ &$E$     \\
   & [kpc]  &[km s$^{-1}$]&[km s$^{-1}$]   &[km s$^{-1}$] & [kpc]& [kpc] &   & [degr.]&[Myr]&[kpc km s$^{-1}$]  &[km$^2$ s$^{-2}$] \\\hline
NGC 104    &   7.6  & $    6^{+  8}_{-  4}$& $  191^{+  5}_{-  5}$& $  196^{+  5}_{-  5}$& $  7.7^{+0.1}_{-0.1}$&
$ 5.51^{+0.26}_{-0.21}$& $0.16^{+0.02}_{-0.02}$& $  28^{+   1}_{-   0}$& $ 116^{+   4}_{-   3}$& $   1323^{+  34}_{-  33}$& $   -126364^{+   1301}_{-    933}$\\
NGC 288    &  12.3  & $    4^{+  1}_{-  1}$& $  -45^{+ 19}_{- 16}$& $   68^{+ 11}_{- 10}$& $ 12.4^{+0.4}_{-0.5}$& $ 1.44^{+0.55}_{-0.58}$& $0.79^{+0.08}_{-0.07}$& $ 121^{+   5}_{-  11}$& $ 142^{+   7}_{-   5}$& $   -374^{+ 160}_{- 136}$& $   -115903^{+   1953}_{-   2171}$\\
NGC 362    &   9.7  & $  133^{+  9}_{- 17}$& $    0^{+ 13}_{-  9}$& $  149^{+  8}_{- 14}$& $ 11.9^{+0.5}_{-0.2}$& $ 0.08^{+0.35}_{-0.00}$& $0.99^{+0.00}_{-0.06}$& $  90^{+  11}_{-  20}$& $ 130^{+   7}_{-   4}$& $      3^{+  95}_{-  65}$& $   -119870^{+   2494}_{-   1943}$\\
Whiting 1    &  35.2  & $ -236^{+ 14}_{- 10}$& $   82^{+  8}_{- 15}$&$250^{+  9}_{- 15}$& $ 79.0^{+12.5}_{-11.5}$& $22.25^{+1.15}_{-2.03}$& $0.56^{+0.05}_{-0.04}$& $  75^{+   2}_{-   2}$& $1406^{+ 243}_{- 231}$& $   1893^{+ 224}_{- 323}$& $    -38160^{+   3070}_{-   3664}$\\
NGC 1261   &  18.3  & $  -98^{+  8}_{-  6}$& $-22^{+  6}_{-  4}$& $  121^{+  6}_{-  8}$& $ 21.3^{+0.8}_{-0.7}$& $ 0.84^{+0.22}_{-0.23}$& $0.92^{+0.02}_{-0.01}$& $ 122^{+   6}_{-   8}$& $ 248^{+   9}_{-  11}$& $   -289^{+  85}_{-  61}$& $    -90868^{+   1747}_{-   1625}$\\
Pal 1     &  17.5  & $   43^{+  7}_{-  3}$& $  215^{+  1}_{-  3}$& $  221^{+  1}_{-  2}$& $ 19.4^{+0.8}_{-0.3}$& $14.89^{+0.37}_{-0.38}$& $0.13^{+0.02}_{-0.01}$& $  15^{+   0}_{-   1}$& $ 358^{+  13}_{-   5}$& $   3691^{+  90}_{-  48}$& $    -77319^{+   1282}_{-    600}$\\
E 1       & 120.3  & $  -21^{+ 35}_{- 49}$& $   -6^{+ 52}_{- 47}$& $   60^{+ 56}_{-  0}$& $129.4^{+26.7}_{-7.0}$& $ 5.20^{+31.55}_{-0.00}$& $0.92^{+0.00}_{-0.35}$& $ 100^{+  21}_{-  34}$& $2140^{+ 923}_{-   0}$& $   -485^{+4161}_{-3811}$& $    -29985^{+   5096}_{-      0}$\\
Eridanus    &  89.8  & $  -80^{+ 17}_{-  9}$& $   22^{+ 11}_{- 21}$& $  163^{+  8}_{- 12}$& $159.2^{+12.1}_{-15.7}$& $13.85^{+5.37}_{-4.11}$& $0.84^{+0.04}_{-0.06}$& $  74^{+  14}_{-   6}$& $2900^{+ 321}_{- 366}$& $   1567^{+ 793}_{-1494}$& $    -25451^{+   1362}_{-   1878}$\\
Pal 2     &  34.3  & $ -107^{+  4}_{-  3}$& $   22^{+ 17}_{- 10}$& $  110^{+  4}_{-  2}$& $ 39.7^{+0.6}_{-2.2}$& $ 1.35^{+1.13}_{-0.60}$& $0.93^{+0.03}_{-0.05}$& $  14^{+  15}_{-   8}$& $ 492^{+  11}_{-  29}$& $    754^{+ 548}_{- 344}$& $    -64910^{+    689}_{-   2083}$\\
NGC 1851   &  16.8  & $  104^{+  1}_{-  3}$& $   -4^{+  5}_{-  3}$& $  132^{+  4}_{-  4}$& $ 19.9^{+0.5}_{-0.5}$& $ 0.17^{+0.12}_{-0.06}$& $0.98^{+0.01}_{-0.01}$& $  97^{+   5}_{-   7}$& $ 228^{+   5}_{-   8}$& $    -66^{+  72}_{-  52}$& $    -94603^{+    985}_{-   1354}$\\
NGC 1904   &  19.2  & $   45^{+  5}_{-  4}$& $    9^{+  4}_{-  9}$& $46^{+  5}_{-  3}$& $ 19.9^{+0.4}_{-0.3}$& $ 0.24^{+0.21}_{-0.17}$& $0.98^{+0.01}_{-0.02}$& $  68^{+  24}_{-   9}$& $ 220^{+   6}_{-   4}$& $    158^{+  68}_{- 172}$& $    -95806^{+   1326}_{-    563}$\\
NGC 2298   &  15.2  & $  -88^{+  7}_{-  8}$& $  -14^{+  6}_{-  9}$& $  112^{+  9}_{-  5}$& $ 16.9^{+0.4}_{-0.4}$& $ 0.51^{+0.25}_{-0.24}$& $0.94^{+0.03}_{-0.03}$& $ 105^{+   8}_{-   7}$& $ 188^{+   6}_{-   3}$& $   -212^{+  91}_{- 129}$& $   -103115^{+   1523}_{-    729}$\\
NGC 2419   &  96.0  & $   -5^{+  7}_{-  2}$& $   46^{+  8}_{- 20}$& $   73^{+  5}_{- 16}$& $ 97.8^{+4.3}_{-2.9}$& $18.02^{+2.06}_{-4.65}$& $0.69^{+0.07}_{-0.03}$& $  52^{+  16}_{-  10}$& $1690^{+  86}_{-  98}$& $4018^{+ 778}_{-1757}$& $    -34413^{+    953}_{-   1094}$\\
Pyxis     &  38.6  & $ -243^{+  4}_{-  4}$& $  -22^{+  8}_{-  5}$& $  292^{+  6}_{-  7}$& $173.6^{+18.1}_{-19.5}$& $18.60^{+0.85}_{-1.28}$& $0.81^{+0.01}_{-0.01}$& $  97^{+   1}_{-   2}$& $3310^{+ 442}_{- 466}$& $   -861^{+ 284}_{- 193}$& $    -23684^{+   1548}_{-   2006}$\\
NGC 2808   &  11.6  & $ -159^{+  1}_{-  1}$& $   36^{+  3}_{-  3}$& $  166^{+  1}_{-  1}$& $14.9^{+0.3}_{-0.3}$& $ 0.90^{+0.08}_{-0.07}$& $0.89^{+0.01}_{-0.02}$& $  11^{+   2}_{-   1}$& $ 162^{+   4}_{-   2}$& $    410^{+  31}_{-  29}$& $   -110129^{+   1092}_{-   1052}$\\
E 3       &   9.2  & $   42^{+ 17}_{-  7}$&$248^{+ 12}_{-  9}$& $271^{+ 13}_{-  8}$& $ 12.4^{+1.6}_{-0.8}$& $ 9.05^{+0.30}_{-0.45}$& $0.15^{+0.07}_{-0.03}$& $  28^{+   2}_{-   1}$& $ 214^{+  17}_{-   9}$& $2184^{+ 135}_{- 131}$& $    -99850^{+   3920}_{-   2651}$\\
Pal 3     &  98.2  & $ -146^{+ 26}_{- 16}$& $   56^{+ 23}_{- 32}$& $  168^{+ 17}_{- 26}$& $148.5^{+49.3}_{-29.3}$& $68.08^{+16.19}_{-20.09}$& $0.37^{+0.13}_{-0.05}$& $  74^{+   9}_{-   7}$& $3742^{+1066}_{- 925}$& $   4168^{+1807}_{-2405}$& $    -22356^{+   3501}_{-   3779}$\\
NGC 3201   &   9.0  & $ -113^{+  6}_{-  9}$& $ -297^{+  6}_{-  6}$& $  351^{+  6}_{-  5}$& $ 24.9^{+2.0}_{-1.6}$& $ 8.29^{+0.20}_{-0.23}$& $0.50^{+0.03}_{-0.02}$& $ 152^{+   1}_{-   1}$& $ 354^{+  27}_{-  22}$& $  -2671^{+  92}_{-  92}$& $    -77493^{+   2730}_{-   2444}$\\
Pal 4      & 104.1  & $   -2^{+ 23}_{- 19}$& $    2^{+ 30}_{- 12}$& $   49^{+ 17}_{-  0}$& $108.7^{+5.7}_{-6.6}$& $ 4.10^{+8.74}_{-1.74}$& $0.93^{+0.03}_{-0.15}$& $  88^{+  10}_{-  14}$& $1712^{+ 139}_{-  65}$& $91^{+1166}_{- 462}$& $    -33868^{+   1358}_{-    807}$\\
Crater    & 147.0  & $  -88^{+ 45}_{- 50}$& $  -26^{+ 88}_{- 63}$& $  106^{+ 87}_{-  6}$& $149.9^{+340.5}_{-45.9}$& $71.70^{+74.52}_{-14.77}$& $0.35^{+0.30}_{-0.08}$& $  99^{+  21}_{-  26}$& $3854^{+1158}_{-3208}$& $  -2527^{+8541}_{-6201}$& $    -21992^{+  13587}_{-   1114}$\\
NGC 4147   &  20.8  & $   42^{+  3}_{-  6}$& $    7^{+ 10}_{-  4}$& $  137^{+  2}_{-  3}$& $ 25.5^{+1.0}_{-0.6}$& $ 0.79^{+0.47}_{-0.15}$& $0.94^{+0.01}_{-0.03}$& $  83^{+   3}_{-   7}$& $ 304^{+  13}_{-   7}$& $     73^{+ 104}_{-  36}$& $    -82528^{+   1719}_{-    888}$\\
NGC 4372   &   7.3  & $   16^{+  6}_{-  8}$& $  132^{+  4}_{-  5}$& $  148^{+  5}_{-  5}$& $  7.3^{+0.2}_{-0.1}$& $ 2.96^{+0.16}_{-0.16}$& $0.42^{+0.03}_{-0.02}$& $  27^{+   1}_{-   1}$& $  98^{+   2}_{-   4}$& $    959^{+  38}_{-  40}$& $   -139945^{+   1706}_{-   1384}$\\
Rup 106    &  18.0  & $ -243^{+  4}_{-  4}$& $   90^{+  8}_{-  6}$& $  261^{+  4}_{-  4}$& $ 36.8^{+2.6}_{-3.4}$& $ 4.48^{+0.45}_{-0.40}$& $0.78^{+0.02}_{-0.02}$& $  46^{+   3}_{-   3}$& $ 480^{+  41}_{-  50}$& $   1585^{+ 162}_{- 165}$& $    -66205^{+   2557}_{-   3720}$\\
NGC 4590   &  10.4  & $ -171^{+  4}_{-  8}$& $  293^{+  3}_{-  8}$& $  340^{+  2}_{-  4}$& $30.5^{+1.4}_{-1.2}$& $ 8.94^{+0.29}_{-0.22}$& $0.55^{+0.01}_{-0.01}$& $  41^{+   2}_{-   1}$& $ 438^{+  22}_{-  19}$& $   2464^{+  71}_{-  75}$& $    -69766^{+   1722}_{-   1527}$\\
NGC 4833   &   7.2  & $  101^{+ 11}_{- 13}$& $   37^{+  8}_{-  7}$& $  116^{+ 11}_{- 12}$& $  8.0^{+0.3}_{-0.3}$& $ 0.64^{+0.16}_{-0.09}$& $0.85^{+0.02}_{-0.03}$& $  39^{+   7}_{-   6}$& $  84^{+   4}_{-   3}$& $    266^{+  57}_{-  55}$& $   -145170^{+   2075}_{-   2171}$\\
NGC 5024   &  19.1  & $  -98^{+  4}_{-  4}$& $  138^{+  7}_{-  6}$& $  184^{+  6}_{-  4}$& $ 23.0^{+1.5}_{-0.7}$& $ 9.15^{+0.98}_{-0.51}$& $0.43^{+0.02}_{-0.02}$& $  75^{+   1}_{-   1}$& $ 344^{+  30}_{-  14}$& $    772^{+  52}_{-  34}$& $    -78916^{+   3067}_{-   1551}$\\
NGC 5053   &  18.1  & $  -91^{+  4}_{-  3}$& $  136^{+  3}_{-  6}$& $  168^{+  3}_{-  6}$& $ 18.1^{+0.2}_{-0.8}$& $10.87^{+0.28}_{-1.01}$& $0.25^{+0.03}_{-0.02}$& $  76^{+   0}_{-   1}$& $ 304^{+   5}_{-  19}$& $    738^{+  25}_{-  38}$& $    -84327^{+    460}_{-   2712}$\\
NGC 5139   &   6.6  & $  -63^{+  5}_{-  5}$& $  -76^{+  5}_{-  4}$& $  131^{+  6}_{-  7}$& $  7.1^{+0.2}_{-0.1}$& $ 1.28^{+0.13}_{-0.14}$& $0.70^{+0.03}_{-0.03}$& $ 135^{+   4}_{-   3}$& $  82^{+   3}_{-   2}$& $   -489^{+  38}_{-  35}$& $   -147265^{+   1628}_{-   1497}$\\
NGC 5272   &  12.2  & $  -38^{+  4}_{-  4}$& $  142^{+  5}_{-  6}$& $  199^{+  3}_{-  4}$& $ 15.9^{+0.7}_{-0.4}$& $ 5.14^{+0.28}_{-0.23}$& $0.51^{+0.02}_{-0.01}$& $  57^{+   1}_{-   1}$& $ 212^{+  10}_{-   6}$& $    988^{+  33}_{-  39}$& $    -98433^{+   2126}_{-   1193}$\\
NGC 5286   &   8.5  & $ -220^{+  2}_{-  1}$& $  -31^{+  8}_{- 11}$& $  223^{+  1}_{-  1}$& $ 13.0^{+0.7}_{-0.5}$& $ 0.54^{+0.20}_{-0.07}$& $0.92^{+0.01}_{-0.03}$& $ 116^{+   8}_{-   6}$& $ 144^{+   7}_{-   5}$& $   -258^{+  66}_{-  87}$& $   -116387^{+   2525}_{-   2131}$\\
NGC 5466   &  16.5  & $  169^{+ 15}_{- 14}$& $ -141^{+ 18}_{- 11}$& $  315^{+  9}_{- 11}$& $ 52.9^{+4.6}_{-6.9}$& $ 5.93^{+0.34}_{-0.81}$& $0.80^{+0.02}_{-0.01}$& $ 108^{+   2}_{-   1}$& $ 736^{+  76}_{- 114}$& $   -816^{+  96}_{-  61}$& $    -53178^{+   2505}_{-   4606}$\\
NGC 5634   &  21.8  & $  -45^{+  6}_{- 15}$& $   41^{+ 10}_{- 19}$& $   66^{+  7}_{-  2}$& $ 22.3^{+0.8}_{-0.7}$& $ 2.29^{+0.43}_{-0.24}$& $0.81^{+0.02}_{-0.03}$& $  70^{+   9}_{-   4}$& $268^{+  12}_{-   9}$& $    383^{+  99}_{- 171}$& $    -87537^{+   1759}_{-   1473}$\\
NGC 5694   &  29.1  & $ -182^{+  6}_{- 11}$& $  -46^{+ 10}_{- 12}$& $  254^{+ 12}_{-  6}$& $ 71.0^{+8.8}_{-4.9}$& $ 2.81^{+0.99}_{-0.75}$& $0.92^{+0.02}_{-0.02}$& $ 134^{+   5}_{-   5}$& $1004^{+ 159}_{-  87}$& $  -1062^{+ 251}_{- 318}$& $    -45021^{+   3362}_{-   2142}$\\
IC 4499   &  15.7  & $ -243^{+  4}_{-  2}$& $  -74^{+  7}_{-  9}$& $  262^{+  2}_{-  2}$& $ 29.9^{+1.7}_{-1.5}$& $ 6.44^{+0.47}_{-0.37}$& $0.65^{+0.02}_{-0.03}$& $ 113^{+   3}_{-   2}$& $ 406^{+  24}_{-  21}$& $  -1056^{+  87}_{- 130}$& $    -72415^{+   2140}_{-   1899}$\\
NGC 5824   &  25.3  & $  -42^{+ 13}_{-  9}$& $  105^{+ 17}_{- 12}$& $  213^{+ 11}_{-  9}$& $ 36.4^{+2.9}_{-2.7}$& $13.54^{+2.28}_{-1.67}$& $0.46^{+0.04}_{-0.05}$& $  58^{+   2}_{-   3}$& $ 576^{+  67}_{-  56}$& $   2343^{+ 412}_{- 300}$& $    -60920^{+   3013}_{-   2792}$\\
Pal 5     &  17.2  & $  -52^{+  2}_{-  3}$& $  138^{+ 31}_{- 16}$& $  148^{+ 29}_{- 15}$& $ 17.6^{+1.0}_{-0.8}$& $ 7.93^{+3.14}_{-1.42}$& $0.38^{+0.07}_{-0.14}$& $  67^{+   3}_{-   2}$& $ 260^{+  44}_{-  20}$& $    962^{+ 249}_{- 180}$& $    -89799^{+   6001}_{-   3264}$\\
NGC 5897   &   7.4  & $   87^{+ 15}_{- 26}$& $   98^{+ 12}_{- 23}$& $  159^{+  9}_{- 19}$& $  8.8^{+0.4}_{-0.4}$& $ 1.94^{+0.35}_{-0.51}$& $0.64^{+0.07}_{-0.05}$& $  60^{+   5}_{-   3}$& $ 106^{+   7}_{-   6}$& $    369^{+  76}_{- 110}$& $   -131407^{+   2594}_{-   3428}$\\
NGC 5904   &   6.3  & $ -290^{+ 12}_{- 11}$& $  126^{+  8}_{-  8}$& $  364^{+  9}_{-  9}$& $ 23.2^{+1.9}_{-2.0}$& $ 2.26^{+0.23}_{-0.19}$& $0.82^{+0.02}_{-0.02}$& $  72^{+   2}_{-   2}$& $ 284^{+  24}_{-  25}$& $    404^{+  28}_{-  35}$& $    -85760^{+   3511}_{-   4080}$\\\hline
 \end{tabular}
   \end{center}
   \end{scriptsize}
  \end{minipage}}

\rotatebox{90}{
		\begin{minipage}{1.5\linewidth}
		\label{t:prop}
		\begin{scriptsize} \begin{center}
{\scriptsize {Table 2:} Orbital properties of the GC. Continued from previous page.}
\medskip
\begin{tabular}{|l|r|r|r|r|r|r|r|r|r|r|r|}\hline
 Name &$d_{GC}$&$\Pi$ &$\Theta$ &$V_{tot}$ &apo   & peri      &ecc&incl. $\theta$    &$T_r$&$L_Z$ &$E$     \\
   & [kpc]  &[km s$^{-1}$]&[km s$^{-1}$]   &[km s$^{-1}$] & [kpc]& [kpc] &   & [degr.]&[Myr]&[kpc km s$^{-1}$]  &[km$^2$ s$^{-2}$] \\\hline
NGC 5927   &   4.8  & $  -51^{+  9}_{- 11}$& $  231^{+  9}_{-  4}$& $  237^{+  8}_{-  3}$& $  5.5^{+0.3}_{-0.2}$& $ 4.13^{+0.25}_{-0.21}$& $0.15^{+0.03}_{-0.03}$& $   9^{+   1}_{-   1}$& $  84^{+   6}_{-   2}$& $   1098^{+  66}_{-  41}$& $   -146878^{+   3102}_{-   1748}$\\
NGC 5946   &   5.2  & $   19^{+ 12}_{- 11}$& $    7^{+  3}_{- 10}$& $  100^{+  4}_{-  4}$& $  5.8^{+0.3}_{-0.3}$& $ 0.06^{+0.08}_{-0.01}$& $0.98^{+0.01}_{-0.03}$& $  86^{+   6}_{-   1}$& $  56^{+   3}_{-   1}$& $     37^{+  14}_{-  52}$& $   -165455^{+   2817}_{-   1113}$\\
ESO 224-8   &  12.6  & $  -44^{+ 18}_{- 24}$& $  257^{+ 17}_{- 16}$& $  261^{+ 17}_{- 13}$& $ 16.8^{+3.4}_{-1.2}$& $11.84^{+0.84}_{-0.98}$& $0.17^{+0.08}_{-0.02}$& $   7^{+   0}_{-   1}$& $ 290^{+  50}_{-  23}$& $   3226^{+ 323}_{- 206}$& $    -85573^{+   6123}_{-   3392}$\\
NGC 5986   &   4.8  & $   64^{+  7}_{- 17}$& $   27^{+  6}_{-  7}$& $   70^{+  8}_{- 15}$& $  5.6^{+0.1}_{-0.6}$& $ 0.20^{+0.10}_{-0.05}$& $0.93^{+0.02}_{-0.04}$& $  63^{+   5}_{-   7}$& $  58^{+   2}_{-   3}$& $    112^{+  22}_{-  29}$& $   -167349^{+   1141}_{-   2618}$\\
FSR 1716   &   4.2  & $   87^{+  9}_{- 19}$& $  162^{+  8}_{-  7}$& $  217^{+  6}_{-  7}$& $  5.2^{+0.0}_{-0.6}$& $ 2.20^{+0.31}_{-0.08}$& $0.41^{+0.00}_{-0.10}$& $  35^{+   2}_{-   2}$& $  68^{+   3}_{-   4}$& $    676^{+  42}_{-  38}$& $   -160568^{+   2517}_{-   2838}$\\
Pal 14    &  68.5  & $  126^{+  9}_{- 15}$& $  -13^{+ 11}_{- 12}$& $  177^{+  9}_{- 11}$& $127.1^{+15.6}_{-10.9}$& $ 1.49^{+2.44}_{-0.47}$& $0.98^{+0.00}_{-0.04}$& $ 130^{+   4}_{-  37}$& $2058^{+ 344}_{- 222}$& $   -597^{+ 537}_{- 587}$& $    -30562^{+   2348}_{-   1823}$\\
BH 184    &   4.4  & $   40^{+  7}_{- 24}$& $  119^{+ 10}_{-  6}$& $  154^{+  7}_{-  5}$& $  4.7^{+0.2}_{-0.3}$& $ 1.65^{+0.21}_{-0.16}$& $0.48^{+0.03}_{-0.05}$& $  36^{+   2}_{-   2}$& $  58^{+   2}_{-   4}$& $    522^{+  49}_{-  42}$& $   -169132^{+   2552}_{-   3715}$\\
NGC 6093   &   3.9  & $   37^{+ 12}_{- 15}$& $   27^{+  6}_{- 16}$& $   78^{+  9}_{- 11}$& $  4.2^{+0.4}_{-0.2}$& $ 0.45^{+0.17}_{-0.26}$& $0.80^{+0.12}_{-0.06}$& $  79^{+   6}_{-   2}$& $  48^{+   1}_{-   3}$& $     50^{+  10}_{-  29}$& $   -173732^{+   1571}_{-   2111}$\\
NGC 6121   &   6.6  & $  -58^{+  1}_{-  2}$& $   47^{+  9}_{-  8}$& $   75^{+  7}_{-  5}$& $  6.8^{+0.1}_{-0.1}$& $ 0.62^{+0.17}_{-0.13}$& $0.83^{+0.03}_{-0.04}$& $   5^{+   1}_{-   0}$& $  74^{+   1}_{-   2}$& $    306^{+  58}_{-  52}$& $   -155030^{+    474}_{-    998}$\\
NGC 6101   &  10.3  & $  -24^{+ 23}_{- 19}$& $ -308^{+  4}_{-  1}$& $  363^{+  2}_{-  4}$& $ 36.3^{+1.6}_{-1.9}$& $10.14^{+0.33}_{-0.43}$& $0.56^{+0.01}_{-0.01}$& $ 143^{+   1}_{-   2}$& $ 532^{+  27}_{-  31}$& $  -2947^{+ 140}_{-  95}$& $    -63205^{+   1597}_{-   2044}$\\
NGC 6144   &   2.5  & $ -180^{+ 60}_{- 25}$& $ -119^{+ 36}_{- 54}$& $  220^{+  3}_{-  2}$& $  3.4^{+0.2}_{-0.2}$& $ 1.56^{+0.25}_{-0.19}$& $0.37^{+0.07}_{-0.08}$& $ 106^{+   6}_{-   4}$& $  50^{+   3}_{-   5}$& $   -146^{+  35}_{-  51}$& $   -174592^{+   1679}_{-   1361}$\\
NGC 6139   &   3.5  & $   -1^{+ 14}_{- 19}$& $   73^{+  4}_{- 11}$& $  150^{+  3}_{-  7}$& $  3.6^{+0.1}_{-0.2}$& $ 0.97^{+0.04}_{-0.16}$& $0.57^{+0.05}_{-0.01}$& $  61^{+   4}_{-   2}$& $  50^{+   0}_{-   4}$& $    237^{+   8}_{-  37}$& $   -178098^{+   1677}_{-   3389}$\\
Terzan 3   &   2.5  & $  -20^{+ 28}_{- 31}$& $  215^{+  6}_{-  9}$& $  234^{+  7}_{-  6}$& $  3.1^{+0.3}_{-0.1}$& $ 2.33^{+0.10}_{-0.24}$& $0.14^{+0.05}_{-0.00}$& $  36^{+   3}_{-   3}$& $  42^{+   6}_{-   2}$& $    472^{+  43}_{-  42}$& $   -176378^{+   3635}_{-   2851}$\\
NGC 6171   &   3.9  & $   -1^{+  2}_{-  2}$& $   98^{+  9}_{-  4}$& $  114^{+  8}_{-  3}$& $  4.0^{+0.2}_{-0.2}$& $ 1.07^{+0.20}_{-0.10}$& $0.57^{+0.03}_{-0.05}$& $  42^{+   2}_{-   2}$& $  50^{+   2}_{-   3}$& $    308^{+  42}_{-  25}$& $   -174517^{+   2798}_{-   2825}$\\
ESO 452-11   &   2.2  & $  -47^{+  7}_{-  9}$& $   22^{+  8}_{- 13}$& $  108^{+  7}_{-  4}$& $  3.0^{+0.2}_{-0.3}$& $ 0.06^{+0.05}_{-0.02}$& $0.96^{+0.01}_{-0.03}$& $  67^{+  11}_{-   6}$& $  28^{+   2}_{-   3}$& $     34^{+  17}_{-  22}$& $   -201978^{+   3413}_{-   3281}$\\
NGC 6205   &   8.7  & $   15^{+  2}_{-  5}$& $  -28^{+  4}_{-  5}$& $   84^{+  6}_{-  3}$& $  8.8^{+0.2}_{-0.2}$& $ 0.97^{+0.10}_{-0.11}$& $0.80^{+0.02}_{-0.02}$& $ 108^{+   3}_{-   2}$& $ 102^{+   3}_{-   2}$& $   -205^{+  30}_{-  35}$& $   -133640^{+    953}_{-   1236}$\\
NGC 6229   &  29.5  & $   32^{+  6}_{-  5}$& $   10^{+  3}_{-  4}$& $   58^{+  5}_{-  5}$& $ 30.6^{+1.3}_{-1.3}$& $ 0.57^{+0.22}_{-0.23}$& $0.96^{+0.02}_{-0.01}$& $  64^{+  11}_{-  12}$& $ 368^{+  18}_{-  18}$& $    227^{+  58}_{-  96}$& $    -75072^{+   1590}_{-   1875}$\\
NGC 6218   &   4.7  & $  -11^{+  4}_{-  4}$& $  128^{+  6}_{-  5}$& $  154^{+  6}_{-  5}$& $  4.9^{+0.2}_{-0.1}$& $ 2.08^{+0.16}_{-0.11}$& $0.40^{+0.03}_{-0.03}$& $  40^{+   1}_{-   2}$& $  64^{+   2}_{-   2}$& $    524^{+  42}_{-  21}$& $   -159771^{+   2293}_{-   1319}$\\
FSR 1735   &   3.2  & $  -96^{+  9}_{-  6}$& $   35^{+ 15}_{- 14}$& $  166^{+  8}_{-  6}$& $  4.2^{+0.3}_{-0.2}$& $ 0.21^{+0.16}_{-0.08}$& $0.90^{+0.04}_{-0.06}$& $  74^{+   6}_{-   6}$& $  42^{+   5}_{-   0}$& $    114^{+  56}_{-  42}$& $   -183293^{+   5174}_{-    920}$\\
NGC 6235   &   4.3  & $  159^{+  7}_{-  4}$& $  213^{+ 20}_{- 25}$& $  269^{+ 18}_{- 19}$& $  7.2^{+0.9}_{-1.1}$& $ 3.13^{+0.34}_{-0.60}$& $0.39^{+0.04}_{-0.01}$& $  50^{+  10}_{-   5}$& $  98^{+  12}_{-  15}$& $    705^{+ 138}_{- 234}$& $   -137634^{+   5829}_{-   8991}$\\
NGC 6254   &   4.5  & $  -91^{+  4}_{-  4}$& $  118^{+  7}_{-  8}$& $  157^{+  5}_{-  5}$& $  4.8^{+0.2}_{-0.2}$& $ 1.78^{+0.15}_{-0.13}$& $0.46^{+0.03}_{-0.04}$& $  43^{+   2}_{-   1}$& $  74^{+   1}_{-   3}$& $    470^{+  31}_{-  29}$& $   -162398^{+   2162}_{-   1569}$\\
NGC 6256   &   2.0  & $  -49^{+ 31}_{- 34}$& $  191^{+  9}_{- 16}$& $  208^{+  5}_{-  6}$& $  2.4^{+0.1}_{-0.1}$& $ 1.53^{+0.48}_{-0.48}$& $0.22^{+0.16}_{-0.12}$& $  25^{+   4}_{-   3}$& $  36^{+   6}_{-   9}$& $    375^{+  75}_{-  74}$& $   -196767^{+   6612}_{-   6208}$\\
Pal 15    &  37.2  & $  155^{+  8}_{-  9}$& $   -1^{+ 13}_{- 10}$& $  162^{+  9}_{-  7}$& $ 52.9^{+3.3}_{-2.9}$& $ 1.30^{+0.97}_{-0.40}$& $0.95^{+0.02}_{-0.03}$& $  92^{+  16}_{-  22}$& $ 700^{+  53}_{-  45}$& $    -38^{+ 403}_{- 334}$& $    -54264^{+   1993}_{-   1884}$\\
NGC 6266   &   2.3  & $   48^{+  8}_{-  6}$& $  127^{+  7}_{- 11}$& $  150^{+  6}_{-  8}$& $  2.7^{+0.3}_{-0.2}$& $ 0.84^{+0.18}_{-0.16}$& $0.53^{+0.05}_{-0.04}$& $  28^{+   2}_{-   1}$& $  36^{+   6}_{-   1}$& $    267^{+  45}_{-  36}$& $   -199401^{+   6501}_{-   4240}$\\
NGC 6273   &   1.5  & $ -245^{+ 83}_{- 32}$& $  -88^{+182}_{- 66}$& $  312^{+  5}_{-  4}$& $  3.5^{+0.4}_{-0.1}$& $ 0.85^{+0.13}_{-0.05}$& $0.61^{+0.02}_{-0.02}$& $  95^{+   6}_{-  15}$& $  44^{+   6}_{-   2}$& $    -40^{+ 122}_{-  49}$& $   -177863^{+   5557}_{-   2069}$\\
NGC 6284   &   6.2  & $   14^{+  3}_{-  2}$& $  -21^{+ 15}_{- 11}$& $  108^{+  3}_{-  5}$& $  6.4^{+0.6}_{-0.6}$& $ 0.51^{+0.10}_{-0.19}$& $0.85^{+0.06}_{-0.03}$& $ 102^{+   6}_{-   9}$& $  76^{+   7}_{-   4}$& $   -120^{+  85}_{-  49}$& $   -151350^{+   4526}_{-   5471}$\\
NGC 6287   &   1.6  & $  302^{+113}_{-300}$& $   77^{+ 56}_{-150}$& $  317^{+  4}_{-  2}$& $  4.4^{+0.8}_{-0.3}$& $ 0.48^{+0.04}_{-0.04}$& $0.81^{+0.03}_{-0.02}$& $  85^{+   5}_{-   6}$& $  52^{+   8}_{-   3}$& $     40^{+  48}_{-  43}$& $   -170400^{+   7687}_{-   3287}$\\
NGC 6293   &   1.6  & $ -162^{+ 64}_{- 31}$& $  -68^{+129}_{- 36}$& $  232^{+  7}_{-  6}$& $  3.2^{+0.5}_{-0.7}$& $ 0.13^{+0.31}_{-0.13}$& $0.92^{+0.09}_{-0.20}$& $ 118^{+  13}_{-  34}$& $  34^{+   5}_{-   3}$& $    -60^{+  93}_{-  56}$& $   -197379^{+   7636}_{-   5832}$\\
NGC 6304   &   2.3  & $   75^{+  6}_{-  7}$& $  189^{+  5}_{-  4}$& $  217^{+  3}_{-  3}$& $  3.1^{+0.4}_{-0.3}$& $ 1.58^{+0.26}_{-0.24}$& $0.32^{+0.03}_{-0.02}$& $  22^{+   1}_{-   1}$& $  48^{+   6}_{-   4}$& $    422^{+  67}_{-  54}$& $   -187732^{+   6888}_{-   6367}$\\
NGC 6316   &   3.0  & $  103^{+  7}_{-  5}$& $   79^{+ 15}_{- 23}$& $  159^{+ 12}_{- 12}$& $  3.9^{+0.6}_{-0.6}$& $ 0.72^{+0.14}_{-0.30}$& $0.69^{+0.10}_{-0.03}$& $  37^{+   8}_{-   5}$& $  46^{+   7}_{-   6}$& $    225^{+  61}_{-  93}$& $   -182890^{+   7119}_{-  10374}$\\
NGC 6341   &   9.9  & $   48^{+  4}_{-  2}$& $   11^{+  4}_{-  3}$& $  110^{+  6}_{-  4}$& $ 10.8^{+0.3}_{-0.2}$& $ 0.45^{+0.13}_{-0.08}$& $0.92^{+0.01}_{-0.02}$& $  81^{+   2}_{-   4}$& $ 126^{+   4}_{-   3}$& $     99^{+  34}_{-  24}$& $   -124522^{+   1736}_{-    972}$\\
NGC 6325   &   1.4  & $  -65^{+  7}_{- 21}$& $ -168^{+ 23}_{- 23}$& $  195^{+ 22}_{- 15}$& $  1.4^{+0.3}_{-0.2}$& $ 1.04^{+0.17}_{-0.29}$& $0.14^{+0.15}_{-0.03}$& $ 123^{+   7}_{-  13}$& $  16^{+  14}_{-   0}$& $   -143^{+  68}_{-  55}$& $   -212303^{+   7398}_{-   7554}$\\
NGC 6333   &   1.8  & $   55^{+ 69}_{-130}$& $  351^{+  9}_{- 33}$& $  362^{+  2}_{-  4}$& $  6.4^{+0.4}_{-0.1}$& $ 0.88^{+0.18}_{-0.08}$& $0.76^{+0.02}_{-0.04}$& $  63^{+   2}_{-   3}$& $  74^{+   6}_{-   1}$& $    284^{+  50}_{-  27}$& $   -151776^{+   3572}_{-   1122}$\\
NGC 6342   &   1.6  & $ -102^{+ 15}_{- 28}$& $  125^{+ 14}_{- 31}$& $  163^{+  5}_{-  3}$& $  1.8^{+0.3}_{-0.1}$& $ 0.63^{+0.11}_{-0.17}$& $0.47^{+0.15}_{-0.07}$& $  64^{+   2}_{-   2}$& $  30^{+   3}_{-   3}$& $    100^{+   9}_{-  10}$& $   -209213^{+   3767}_{-   1126}$\\
NGC 6356   &   7.8  & $   47^{+  8}_{-  7}$& $  117^{+ 21}_{- 23}$& $  167^{+ 14}_{- 14}$& $  8.5^{+0.9}_{-0.7}$& $ 2.96^{+0.78}_{-0.75}$& $0.48^{+0.10}_{-0.08}$& $  41^{+   5}_{-   3}$& $ 114^{+  13}_{-   9}$& $    846^{+ 177}_{- 186}$& $   -131325^{+   5976}_{-   5365}$\\
NGC 6355   &   0.9  & $ -218^{+302}_{- 47}$& $  -89^{+227}_{- 34}$& $  273^{+  5}_{-  4}$& $  1.4^{+0.7}_{-0.0}$& $ 0.64^{+0.05}_{-0.40}$& $0.38^{+0.41}_{-0.01}$& $  97^{+   7}_{-  22}$& $  20^{+   5}_{-   2}$& $    -29^{+  72}_{-  35}$& $   -216289^{+   9034}_{-   3431}$\\
NGC 6352   &   3.6  & $   44^{+ 10}_{- 10}$& $  226^{+  6}_{-  6}$& $  230^{+  6}_{-  6}$& $  4.2^{+0.3}_{-0.3}$& $ 3.19^{+0.15}_{-0.19}$& $0.13^{+0.03}_{-0.02}$& $  11^{+   1}_{-   1}$& $  70^{+   2}_{-   5}$& $    802^{+  49}_{-  54}$& $   -163337^{+   3190}_{-   3394}$\\
IC 1257   &  19.2  & $  -51^{+  9}_{-  7}$& $  -23^{+ 11}_{- 18}$& $   58^{+ 10}_{-  3}$& $ 20.0^{+0.6}_{-2.0}$& $ 0.75^{+0.74}_{-0.33}$& $0.93^{+0.03}_{-0.08}$& $ 157^{+   4}_{-  13}$& $ 224^{+   8}_{-  20}$& $   -407^{+ 200}_{- 272}$& $    -95179^{+   1696}_{-   4162}$\\
\hline
 \end{tabular}
   \end{center}
   \end{scriptsize}
  \end{minipage}}

\rotatebox{90}{
		\begin{minipage}{1.5\linewidth}
	     \label{t:prop}
		\begin{scriptsize} \begin{center}
{\scriptsize {Table 2:} Orbital properties of the GC. Continued from previous page.}
\medskip
\begin{tabular}{|l|r|r|r|r|r|r|r|r|r|r|r|}\hline
 Name &$d_{GC}$&$\Pi$ &$\Theta$ &$V_{tot}$ &apo   & peri      &ecc&incl. $\theta$    &$T_r$&$L_Z$ &$E$     \\
   & [kpc]  &[km s$^{-1}$]&[km s$^{-1}$]   &[km s$^{-1}$] & [kpc]& [kpc] &   & [degr.]&[Myr]&[kpc km s$^{-1}$]  &[km$^2$ s$^{-2}$] \\\hline
Terzan 2   &   0.8  & $ -104^{+ 63}_{- 25}$& $  -74^{+ 26}_{- 36}$& $  137^{+  3}_{-  1}$& $  1.0^{+0.3}_{-0.3}$& $ 0.13^{+0.04}_{-0.04}$& $0.76^{+0.10}_{-0.16}$& $ 156^{+   3}_{-   9}$& $  10^{+   4}_{-   3}$& $    -56^{+  11}_{-  10}$& $   -252725^{+  13021}_{-  13814}$\\
NGC 6366   &   5.3  & $   94^{+  2}_{-  2}$& $  135^{+  2}_{-  3}$& $  176^{+  2}_{-  3}$& $  5.9^{+0.1}_{-0.1}$& $ 2.24^{+0.06}_{-0.08}$& $0.45^{+0.01}_{-0.01}$& $  32^{+   1}_{-   1}$& $  76^{+   2}_{-   2}$& $    711^{+  20}_{-  26}$& $   -152836^{+   1168}_{-   1273}$\\
Terzan 4   &   0.9  & $   13^{+ 11}_{- 25}$& $   61^{+  9}_{-  5}$& $  118^{+  8}_{-  3}$& $  0.9^{+0.3}_{-0.2}$& $ 0.18^{+0.02}_{-0.03}$& $0.68^{+0.05}_{-0.05}$& $  58^{+   4}_{-   4}$& $  10^{+   4}_{-   2}$& $     55^{+  21}_{-  13}$& $   -250150^{+  13896}_{-  14886}$\\
BH 229    &   1.4  & $  -54^{+ 11}_{-  8}$& $    7^{+ 12}_{- 19}$& $  250^{+ 16}_{- 10}$& $  2.4^{+0.5}_{-0.7}$& $ 0.24^{+0.42}_{-0.30}$& $0.82^{+0.18}_{-0.24}$& $  88^{+   4}_{-   3}$& $  26^{+   7}_{-   9}$& $     10^{+  14}_{-  17}$& $   -205120^{+  14227}_{-  22643}$\\
FSR 1758   &   3.4  & $   45^{+ 27}_{- 41}$& $ -343^{+  5}_{-  3}$& $  396^{+  4}_{-  5}$& $ 12.0^{+1.2}_{-1.3}$& $ 3.31^{+0.37}_{-0.39}$& $0.57^{+0.01}_{-0.02}$& $ 148^{+   1}_{-   1}$& $ 152^{+  16}_{-  17}$& $  -1133^{+ 128}_{- 121}$& $   -115174^{+   5171}_{-   6059}$\\
NGC 6362   &   5.2  & $   18^{+ 16}_{- 19}$& $  124^{+  8}_{-  9}$& $  160^{+  6}_{-  4}$& $  5.4^{+0.2}_{-0.1}$& $ 2.48^{+0.16}_{-0.20}$& $0.37^{+0.03}_{-0.02}$& $  45^{+   3}_{-   3}$& $  72^{+   3}_{-   3}$& $    585^{+  41}_{-  36}$& $   -153270^{+   2099}_{-   1855}$\\
Liller 1   &   0.8  & $   80^{+ 27}_{- 45}$& $  -70^{+ 37}_{- 32}$& $  109^{+ 21}_{- 18}$& $  0.8^{+0.3}_{-0.0}$& $ 0.12^{+0.11}_{-0.06}$& $0.75^{+0.12}_{-0.11}$& $ 160^{+  15}_{-  33}$& $  10^{+   3}_{-   2}$& $    -54^{+  30}_{-  47}$& $   -261115^{+  16395}_{-   4011}$\\
NGC 6380   &   2.1  & $  -82^{+  8}_{-  8}$& $  -27^{+ 18}_{-  9}$& $   87^{+  8}_{-  9}$& $  2.4^{+0.2}_{-0.3}$& $ 0.10^{+0.05}_{-0.06}$& $0.92^{+0.04}_{-0.03}$& $ 153^{+  12}_{-  38}$& $  28^{+   3}_{-   4}$& $    -54^{+  37}_{-  24}$& $   -212458^{+   4778}_{-   5472}$\\
Terzan 1   &   2.7  & $  -72^{+  2}_{-  2}$& $   98^{+  9}_{-  9}$& $  121^{+  8}_{-  7}$& $  2.8^{+0.2}_{-0.2}$& $ 0.67^{+0.13}_{-0.11}$& $0.62^{+0.04}_{-0.05}$& $   3^{+   0}_{-   0}$& $  38^{+   2}_{-   3}$& $    259^{+  34}_{-  33}$& $   -200171^{+   3293}_{-   3450}$\\
Pismis 26   &   1.9  & $   44^{+ 18}_{- 27}$& $  238^{+  6}_{-  6}$& $  296^{+  6}_{-  3}$& $  3.3^{+0.5}_{-0.4}$& $ 1.76^{+0.21}_{-0.18}$& $0.30^{+0.02}_{-0.01}$& $  38^{+   1}_{-   1}$& $  52^{+  11}_{-   2}$& $    429^{+  65}_{-  52}$& $   -178784^{+   6979}_{-   5703}$\\
NGC 6388   &   3.9  & $  -29^{+  8}_{- 15}$& $  -92^{+  7}_{-  3}$& $   98^{+  5}_{-  5}$& $  4.2^{+0.2}_{-0.3}$& $ 1.00^{+0.12}_{-0.24}$& $0.61^{+0.07}_{-0.02}$& $ 155^{+   2}_{-   5}$& $  52^{+   3}_{-   4}$& $   -338^{+  57}_{-  27}$& $   -178537^{+   3036}_{-   5190}$\\
NGC 6402   &   4.0  & $  -18^{+ 13}_{- 15}$& $   46^{+  5}_{-  5}$& $   53^{+  8}_{-  4}$& $  4.7^{+0.1}_{-0.3}$& $ 0.27^{+0.12}_{-0.02}$& $0.89^{+0.01}_{-0.05}$& $  46^{+   5}_{-   5}$& $  52^{+   1}_{-   4}$& $    149^{+  17}_{-  19}$& $   -177531^{+   1637}_{-   2128}$\\
NGC 6401   &   0.8  & $  257^{+  8}_{- 76}$& $   34^{+114}_{-132}$& $  285^{+  2}_{-  2}$& $  2.0^{+0.6}_{-0.4}$& $ 0.06^{+0.36}_{-0.05}$& $0.94^{+0.02}_{-0.25}$& $  77^{+  33}_{-  38}$& $  18^{+  12}_{-   3}$& $     19^{+ 119}_{-  88}$& $   -221130^{+  14747}_{-   3396}$\\
NGC 6397   &   6.1  & $   41^{+  3}_{-  6}$& $  118^{+  4}_{-  8}$& $  181^{+  5}_{-  7}$& $  6.5^{+0.1}_{-0.2}$& $ 2.57^{+0.14}_{-0.26}$& $0.43^{+0.04}_{-0.02}$& $  47^{+   3}_{-   1}$& $  84^{+   2}_{-   4}$& $    721^{+  16}_{-  59}$& $   -145511^{+   1125}_{-   2578}$\\
Pal 6     &   1.3  & $ -199^{+  4}_{-  4}$& $  -10^{+ 25}_{- 21}$& $  270^{+ 10}_{-  3}$& $  2.8^{+0.8}_{-0.8}$& $ 0.08^{+0.08}_{-0.07}$& $0.95^{+0.07}_{-0.11}$& $  93^{+   6}_{-   6}$& $  28^{+   8}_{-   7}$& $    -13^{+  28}_{-  36}$& $   -202849^{+  14610}_{-  15486}$\\
NGC 6426   &  14.4  & $ -111^{+ 12}_{- 12}$& $   94^{+  6}_{-  9}$& $  149^{+ 12}_{- 14}$& $ 16.7^{+0.8}_{-0.7}$& $ 3.28^{+0.26}_{-0.41}$& $0.67^{+0.03}_{-0.01}$& $  26^{+   2}_{-   2}$& $ 204^{+  11}_{-  12}$& $   1240^{+  95}_{- 136}$& $   -100200^{+   2396}_{-   2590}$\\
Djorg 1    &   1.7  & $ -291^{+ 48}_{- 41}$& $  294^{+ 39}_{- 48}$& $  414^{+  8}_{- 10}$& $  8.6^{+2.2}_{-1.6}$& $ 1.06^{+0.18}_{-0.15}$& $0.78^{+0.02}_{-0.02}$& $  19^{+   3}_{-   4}$& $  94^{+  24}_{-  16}$& $    487^{+  97}_{-  82}$& $   -140346^{+  13303}_{-  12077}$\\
Terzan 5   &   1.8  & $   75^{+  4}_{-  2}$& $   62^{+  7}_{-  5}$& $  102^{+  5}_{-  2}$& $  1.9^{+0.3}_{-0.2}$& $ 0.22^{+0.06}_{-0.05}$& $0.80^{+0.02}_{-0.03}$& $  32^{+   5}_{-   3}$& $  24^{+   3}_{-   4}$& $    108^{+  25}_{-  20}$& $   -220380^{+   6113}_{-   5974}$\\
NGC 6440   &   1.3  & $   94^{+  4}_{- 19}$& $  -22^{+ 29}_{- 28}$& $  104^{+  3}_{-  9}$& $  1.5^{+0.1}_{-0.2}$& $ 0.05^{+0.12}_{-0.02}$& $0.93^{+0.04}_{-0.15}$& $ 104^{+  17}_{-  19}$& $  14^{+   2}_{-   1}$& $    -24^{+  34}_{-  34}$& $   -233545^{+   4082}_{-    591}$\\
NGC 6441   &   4.7  & $   17^{+  4}_{-  7}$& $   99^{+ 10}_{- 18}$& $  103^{+ 10}_{- 17}$& $  4.7^{+0.7}_{-0.6}$& $ 1.43^{+0.26}_{-0.35}$& $0.53^{+0.08}_{-0.03}$& $  18^{+   3}_{-   2}$& $  58^{+   8}_{-   7}$& $    447^{+  90}_{- 107}$& $   -169955^{+   7831}_{-   7536}$\\
Terzan 6   &   1.1  & $ -129^{+ 48}_{- 24}$& $  -79^{+ 11}_{- 28}$& $  153^{+  4}_{-  4}$& $  1.3^{+0.4}_{-0.7}$& $ 0.17^{+0.04}_{-0.04}$& $0.77^{+0.05}_{-0.16}$& $ 164^{+   6}_{-  19}$& $  16^{+   5}_{-   9}$& $    -83^{+  34}_{-  20}$& $   -236580^{+  13730}_{-  37032}$\\
NGC 6453   &   2.0  & $ -109^{+ 16}_{-  7}$& $   20^{+ 32}_{- 18}$& $  178^{+  9}_{-  4}$& $  2.6^{+0.6}_{-0.9}$& $ 0.20^{+0.43}_{-0.03}$& $0.86^{+0.04}_{-0.38}$& $  84^{+   6}_{-   9}$& $  32^{+   8}_{-   8}$& $     38^{+  36}_{-  36}$& $   -201623^{+  10681}_{-  14973}$\\
NGC 6496   &   2.8  & $   28^{+ 23}_{- 31}$& $  264^{+  9}_{- 11}$& $  269^{+ 11}_{- 11}$& $  4.6^{+0.4}_{-0.3}$& $ 2.35^{+0.22}_{-0.15}$& $0.32^{+0.05}_{-0.05}$& $  37^{+   1}_{-   2}$& $  64^{+   4}_{-   5}$& $    581^{+  46}_{-  31}$& $   -161433^{+   3819}_{-   2581}$\\
Terzan 9   &   2.6  & $  -45^{+  4}_{-  2}$& $   52^{+ 14}_{-  4}$& $   84^{+  9}_{-  3}$& $  2.7^{+0.2}_{-0.2}$& $ 0.28^{+0.09}_{-0.05}$& $0.81^{+0.02}_{-0.05}$& $  45^{+   2}_{-   7}$& $  32^{+   2}_{-   2}$& $    132^{+  39}_{-  17}$& $   -205042^{+   3295}_{-   4485}$\\
Djorg 2    &   0.7  & $    8^{+ 99}_{- 77}$& $ -206^{+ 62}_{- 13}$& $  218^{+  4}_{-  4}$& $  0.8^{+0.6}_{-0.2}$& $ 0.50^{+0.30}_{-0.21}$& $0.21^{+0.31}_{-0.10}$& $ 146^{+   6}_{-  10}$& $  14^{+   8}_{-   5}$& $   -126^{+  51}_{-  81}$& $   -246126^{+  22397}_{-  10440}$\\
NGC 6517   &   3.2  & $   58^{+  4}_{- 16}$& $   41^{+ 10}_{-  4}$& $   78^{+  4}_{-  9}$& $  3.7^{+0.2}_{-0.2}$& $ 0.23^{+0.09}_{-0.03}$& $0.88^{+0.02}_{-0.04}$& $  53^{+   2}_{-   8}$& $  42^{+   2}_{-   2}$& $    125^{+  31}_{-  12}$& $   -190163^{+   2504}_{-   2700}$\\
Terzan 10   &   2.1  & $  228^{+ 25}_{- 46}$& $   94^{+ 54}_{- 30}$& $  335^{+  9}_{- 10}$& $  5.3^{+1.0}_{-1.3}$& $ 0.57^{+0.12}_{-0.09}$& $0.81^{+0.06}_{-0.11}$& $  70^{+   5}_{-   9}$& $  70^{+  10}_{-  16}$& $    191^{+  31}_{-  32}$& $   -162093^{+  10123}_{-  15831}$\\
NGC 6522   &   1.1  & $   26^{+  5}_{-  5}$& $  102^{+  7}_{-  8}$& $  206^{+ 11}_{-  6}$& $  1.4^{+0.5}_{-0.2}$& $ 0.42^{+0.35}_{-0.11}$& $0.54^{+0.06}_{-0.12}$& $  59^{+   3}_{-   1}$& $  24^{+   8}_{-   7}$& $    105^{+  54}_{-  27}$& $   -222763^{+  17892}_{-  10688}$\\
NGC 6535   &   4.1  & $  109^{+  5}_{-  7}$& $  -69^{+  5}_{-  8}$& $  135^{+  4}_{-  3}$& $  4.8^{+0.1}_{-0.2}$& $ 0.80^{+0.11}_{-0.11}$& $0.72^{+0.03}_{-0.04}$& $ 163^{+   0}_{-   2}$& $  56^{+   2}_{-   2}$& $   -273^{+  27}_{-  31}$& $   -172105^{+    804}_{-   2255}$\\
NGC 6528   &   0.8  & $ -201^{+217}_{- 66}$& $  103^{+ 62}_{- 38}$& $  230^{+  4}_{-  2}$& $  1.1^{+0.8}_{-0.4}$& $ 0.23^{+0.15}_{-0.11}$& $0.66^{+0.23}_{-0.33}$& $  70^{+   5}_{-   3}$& $  16^{+   7}_{-   6}$& $     53^{+  15}_{-  17}$& $   -238646^{+  17943}_{-  12004}$\\
NGC 6539   &   3.1  & $    1^{+ 16}_{- 16}$& $  118^{+  4}_{-  7}$& $  214^{+  8}_{-  7}$& $  3.5^{+0.1}_{-0.2}$& $ 1.85^{+0.27}_{-0.14}$& $0.31^{+0.04}_{-0.07}$& $  57^{+   2}_{-   2}$& $  62^{+   7}_{-  11}$& $    349^{+  16}_{-  30}$& $   -172417^{+   2008}_{-   2527}$\\
NGC 6540   &   2.5  & $   16^{+  3}_{-  3}$& $  136^{+  5}_{-  6}$& $  150^{+  5}_{-  5}$& $  2.6^{+0.3}_{-0.2}$& $ 1.14^{+0.19}_{-0.12}$& $0.38^{+0.04}_{-0.03}$& $  26^{+   1}_{-   1}$& $  40^{+   4}_{-   3}$& $    331^{+  50}_{-  28}$& $   -198910^{+   6319}_{-   3631}$\\
NGC 6544   &   5.7  & $   10^{+  2}_{-  1}$& $   40^{+  5}_{-  9}$& $   87^{+  3}_{-  6}$& $  6.0^{+0.1}_{-0.2}$& $ 0.45^{+0.08}_{-0.13}$& $0.86^{+0.04}_{-0.02}$& $  63^{+   5}_{-   3}$& $  64^{+   2}_{-   2}$& $    229^{+  27}_{-  52}$& $   -162598^{+   1056}_{-   1837}$\\
NGC 6541   &   2.2  & $  117^{+ 33}_{- 47}$& $  197^{+ 22}_{- 21}$& $  255^{+  7}_{-  3}$& $  3.7^{+0.4}_{-0.2}$& $ 1.29^{+0.23}_{-0.15}$& $0.49^{+0.07}_{-0.08}$& $  41^{+   3}_{-   4}$& $  50^{+   4}_{-   3}$& $    333^{+  34}_{-  15}$& $   -175101^{+   4019}_{-   1532}$\\
ESO 280-06   &  13.3  & $   34^{+  7}_{-  3}$& $   21^{+ 16}_{- 11}$& $   83^{+  9}_{-  1}$& $ 13.7^{+0.6}_{-0.9}$& $ 0.73^{+0.45}_{-0.22}$& $0.90^{+0.03}_{-0.06}$& $  71^{+  10}_{-  14}$& $ 156^{+   8}_{-   9}$& $    257^{+ 191}_{- 135}$& $   -112187^{+   2270}_{-   3282}$\\
NGC 6553   &   3.0  & $   29^{+  5}_{-  3}$& $  249^{+  1}_{-  1}$& $  251^{+  1}_{-  1}$& $  3.9^{+0.2}_{-0.3}$& $ 2.96^{+0.18}_{-0.30}$& $0.13^{+0.02}_{-0.01}$& $   5^{+   1}_{-   0}$& $  62^{+   3}_{-   3}$& $    756^{+  46}_{-  74}$& $   -168744^{+   2830}_{-   4839}$\\
NGC 6558   &   1.2  & $  187^{+  3}_{-  1}$& $   91^{+  6}_{- 13}$& $  209^{+  3}_{-  4}$& $  1.6^{+0.4}_{-0.5}$& $ 0.26^{+0.17}_{-0.08}$& $0.72^{+0.13}_{-0.25}$& $  65^{+  11}_{-   6}$& $  18^{+   5}_{-   4}$& $     79^{+  25}_{-  38}$& $   -219183^{+   7780}_{-  11838}$\\
Pal 7     &   4.5  & $ -102^{+  6}_{-  5}$& $  268^{+  6}_{-  4}$& $  287^{+  5}_{-  3}$& $  7.2^{+0.5}_{-0.2}$& $ 3.78^{+0.24}_{-0.11}$& $0.31^{+0.01}_{-0.01}$& $  10^{+   0}_{-   1}$& $ 100^{+   6}_{-   4}$& $   1184^{+  85}_{-  39}$& $   -138577^{+   4025}_{-   1922}$\\
\hline
 \end{tabular}
   \end{center}
   \end{scriptsize}
  \end{minipage}}

\rotatebox{90}{
		\begin{minipage}{1.5\linewidth}
	     \label{t:prop}
		\begin{scriptsize} \begin{center}
{\scriptsize {Table 2:} Orbital properties of the GC. Continued from previous page.}
\medskip
\begin{tabular}{|l|r|r|r|r|r|r|r|r|r|r|r|}\hline
Name &$d_{GC}$&$\Pi$ &$\Theta$ &$V_{tot}$ &apo   & peri      &ecc&incl. $\theta$    &$T_r$&$L_Z$ &$E$     \\
   & [kpc]  &[km s$^{-1}$]&[km s$^{-1}$]   &[km s$^{-1}$] & [kpc]& [kpc] &   & [degr.]&[Myr]&[kpc km s$^{-1}$]  &[km$^2$ s$^{-2}$] \\\hline
Terzan 12   &   3.3  & $  -94^{+  5}_{-  2}$& $  160^{+  8}_{-  6}$& $  211^{+  6}_{-  5}$& $  4.0^{+0.3}_{-0.3}$& $ 1.87^{+0.15}_{-0.13}$& $0.36^{+0.02}_{-0.03}$& $  31^{+   2}_{-   2}$& $  54^{+   5}_{-   2}$& $    524^{+  45}_{-  40}$& $   -174529^{+   3905}_{-   4020}$\\
NGC 6569   &   2.5  & $  -40^{+  4}_{-  4}$& $  164^{+ 14}_{- 25}$& $  170^{+ 14}_{- 24}$& $  2.6^{+0.6}_{-0.4}$& $ 1.46^{+0.25}_{-0.43}$& $0.28^{+0.15}_{-0.05}$& $  29^{+  10}_{-   7}$& $  46^{+   7}_{-   8}$& $    354^{+  75}_{-  95}$& $   -189983^{+   6899}_{-   7687}$\\
ESO 456-78   &   2.3  & $   62^{+  4}_{-  4}$& $  203^{+  2}_{-  3}$& $  249^{+  4}_{-  3}$& $  3.2^{+0.3}_{-0.3}$& $ 1.81^{+0.28}_{-0.26}$& $0.27^{+0.04}_{-0.02}$& $  32^{+   2}_{-   1}$& $  58^{+   4}_{-   6}$& $    458^{+  64}_{-  59}$& $   -180481^{+   5218}_{-   5189}$\\
NGC 6584   &   6.9  & $  198^{+ 12}_{- 16}$& $  105^{+ 29}_{- 12}$& $  327^{+ 12}_{-  9}$& $ 18.6^{+2.5}_{-1.2}$& $ 1.79^{+0.77}_{-0.27}$& $0.82^{+0.03}_{-0.05}$& $  50^{+   3}_{-   4}$& $ 220^{+  34}_{-  15}$& $    603^{+ 231}_{-  83}$& $    -96417^{+   6158}_{-   3082}$\\
NGC 6624   &   1.2  & $  -23^{+ 44}_{- 26}$& $   60^{+  4}_{- 18}$& $  138^{+  6}_{-  5}$& $  1.7^{+0.3}_{-0.1}$& $ 0.08^{+0.10}_{-0.05}$& $0.91^{+0.05}_{-0.09}$& $  74^{+   2}_{-   4}$& $  20^{+   4}_{-   4}$& $     32^{+  16}_{-  11}$& $   -227013^{+   7470}_{-   2603}$\\
NGC 6626   &   3.1  & $  -27^{+  4}_{-  2}$& $   61^{+ 12}_{-  6}$& $  113^{+  7}_{-  4}$& $  3.2^{+0.3}_{-0.3}$& $ 0.49^{+0.11}_{-0.14}$& $0.73^{+0.08}_{-0.06}$& $  58^{+   3}_{-   5}$& $  44^{+   2}_{-   9}$& $    190^{+  32}_{-  28}$& $   -191218^{+   2777}_{-   6026}$\\
NGC 6638   &   2.2  & $   64^{+  9}_{- 12}$& $   10^{+ 13}_{- 16}$& $   70^{+  6}_{-  5}$& $  2.7^{+0.3}_{-0.2}$& $ 0.04^{+0.06}_{-0.01}$& $0.97^{+0.01}_{-0.04}$& $  81^{+  13}_{-  10}$& $  24^{+   7}_{-   3}$& $     20^{+  18}_{-  30}$& $   -206839^{+   8133}_{-   4599}$\\
NGC 6637   &   1.7  & $   43^{+ 20}_{-113}$& $   89^{+  9}_{- 58}$& $  128^{+  7}_{-  6}$& $  2.4^{+0.2}_{-0.2}$& $ 0.12^{+0.10}_{-0.13}$& $0.91^{+0.10}_{-0.09}$& $  75^{+  11}_{-   2}$& $  22^{+   2}_{-   0}$& $     47^{+  12}_{-  37}$& $   -211371^{+   4131}_{-   1668}$\\
NGC 6642   &   1.7  & $  112^{+  6}_{-  9}$& $   29^{+ 24}_{- 36}$& $  126^{+  8}_{-  5}$& $  2.2^{+0.2}_{-0.1}$& $ 0.09^{+0.11}_{-0.05}$& $0.92^{+0.04}_{-0.05}$& $  43^{+  58}_{-  18}$& $  24^{+   4}_{-   5}$& $     42^{+  42}_{-  51}$& $   -215290^{+   4209}_{-   2746}$\\
NGC 6652   &   2.1  & $  -55^{+ 12}_{-  7}$& $   10^{+ 16}_{- 23}$& $  175^{+  9}_{-  7}$& $  3.6^{+0.5}_{-0.4}$& $ 0.03^{+0.08}_{-0.01}$& $0.98^{+0.01}_{-0.02}$& $  81^{+  11}_{-   6}$& $  32^{+   5}_{-   3}$& $     10^{+  21}_{-  15}$& $   -192830^{+   7171}_{-   3378}$\\
NGC 6656   &   5.1  & $  177^{+  2}_{-  1}$& $  199^{+  1}_{-  1}$& $  305^{+  8}_{-  4}$& $  9.8^{+0.4}_{-0.3}$& $ 2.99^{+0.09}_{-0.11}$& $0.53^{+0.01}_{-0.00}$& $  34^{+   3}_{-   2}$& $ 124^{+   6}_{-   3}$& $   1013^{+  26}_{-  33}$& $   -125755^{+   2366}_{-   1764}$\\
Pal 8     &   4.0  & $  -17^{+ 13}_{- 18}$& $   78^{+ 11}_{- 11}$& $   85^{+ 15}_{- 11}$& $  4.2^{+0.2}_{-0.2}$& $ 0.87^{+0.11}_{-0.17}$& $0.66^{+0.05}_{-0.03}$& $  29^{+   3}_{-   2}$& $  52^{+   1}_{-   4}$& $    295^{+  39}_{-  47}$& $   -178566^{+   2806}_{-   3432}$\\
NGC 6681   &   2.2  & $  223^{+ 16}_{- 45}$& $    9^{+ 87}_{- 45}$& $  287^{+  4}_{-  6}$& $  5.0^{+0.6}_{-0.5}$& $ 0.48^{+0.27}_{-0.18}$& $0.82^{+0.07}_{-0.10}$& $  88^{+  13}_{-   8}$& $  56^{+   4}_{-   3}$& $      8^{+  45}_{-  43}$& $   -163960^{+   3687}_{-   3159}$\\
NGC 6712   &   3.6  & $  141^{+  3}_{-  4}$& $    5^{+ 10}_{-  8}$& $  213^{+  5}_{-  6}$& $  5.6^{+0.3}_{-0.4}$& $ 0.07^{+0.05}_{-0.03}$& $0.98^{+0.01}_{-0.02}$& $  88^{+   3}_{-   4}$& $  58^{+   4}_{-   4}$& $     18^{+  35}_{-  27}$& $   -168037^{+   3989}_{-   3245}$\\
NGC 6715   &  18.4  & $  231^{+  3}_{-  3}$& $   48^{+ 13}_{- 16}$& $  311^{+  5}_{- 12}$& $ 51.8^{+8.7}_{-8.2}$& $14.41^{+0.76}_{-1.06}$& $0.56^{+0.04}_{-0.03}$& $  80^{+   3}_{-   3}$& $ 826^{+ 153}_{- 143}$& $    836^{+ 231}_{- 266}$& $    -50574^{+   3996}_{-   4779}$\\
NGC 6717   &   2.4  & $   -1^{+ 27}_{- 18}$& $  111^{+  5}_{-  7}$& $  114^{+  7}_{-  5}$& $  2.7^{+0.5}_{-0.2}$& $ 0.64^{+0.15}_{-0.07}$& $0.62^{+0.05}_{-0.08}$& $  37^{+   5}_{-   5}$& $  32^{+   7}_{-   0}$& $    220^{+  40}_{-  23}$& $   -197801^{+   4895}_{-   2473}$\\
NGC 6723   &   2.5  & $  105^{+ 42}_{- 15}$& $  180^{+ 17}_{- 48}$& $  211^{+  2}_{-  4}$& $  3.1^{+0.1}_{-0.2}$& $ 1.68^{+0.07}_{-0.10}$& $0.30^{+0.02}_{-0.03}$& $  82^{+   6}_{-   7}$& $  42^{+   1}_{-   4}$& $     73^{+  57}_{-  53}$& $   -176770^{+   1834}_{-   3157}$\\
NGC 6749   &   5.0  & $  -20^{+ 10}_{- 11}$& $  107^{+  5}_{- 10}$& $  109^{+  5}_{-  8}$& $  5.0^{+0.3}_{-0.1}$& $ 1.47^{+0.13}_{-0.17}$& $0.55^{+0.04}_{-0.03}$& $   3^{+   0}_{-   0}$& $  62^{+   4}_{-   1}$& $    532^{+  39}_{-  44}$& $   -168249^{+   3149}_{-   1759}$\\
NGC 6752   &   5.4  & $  -25^{+  3}_{-  4}$& $  177^{+  2}_{-  5}$& $  188^{+  2}_{-  5}$& $  5.6^{+0.1}_{-0.1}$& $ 3.46^{+0.07}_{-0.18}$& $0.24^{+0.02}_{-0.01}$& $  24^{+   1}_{-   1}$& $  80^{+   3}_{-   1}$& $    903^{+  21}_{-  39}$& $   -148003^{+   1066}_{-   1778}$\\
NGC 6760   &   5.2  & $  101^{+  9}_{-  9}$& $  132^{+  6}_{-  9}$& $  167^{+  6}_{-  7}$& $  5.9^{+0.1}_{-0.4}$& $ 1.94^{+0.11}_{-0.23}$& $0.50^{+0.04}_{-0.02}$& $   7^{+   1}_{-   0}$& $  74^{+   0}_{-   5}$& $    684^{+  24}_{-  70}$& $   -157039^{+   1094}_{-   3811}$\\
NGC 6779   &   9.9  & $  153^{+  1}_{-  2}$& $  -30^{+  7}_{-  4}$& $  192^{+  4}_{-  4}$& $ 13.2^{+0.7}_{-0.4}$& $ 0.71^{+0.21}_{-0.21}$& $0.90^{+0.03}_{-0.03}$& $ 109^{+   2}_{-   4}$& $ 150^{+   8}_{-   8}$& $   -294^{+  70}_{-  46}$& $   -114425^{+   2707}_{-   2516}$\\
Terzan 7   &  16.7  & $  264^{+  5}_{-  5}$& $   49^{+ 12}_{- 21}$& $  333^{+  6}_{-  8}$& $ 58.0^{+5.5}_{-9.8}$& $14.34^{+0.48}_{-1.23}$& $0.60^{+0.03}_{-0.04}$& $  82^{+   3}_{-   2}$& $ 924^{+  96}_{- 170}$& $    717^{+ 137}_{- 311}$& $    -47611^{+   2688}_{-   5710}$\\
Pal 10    &   7.7  & $  -89^{+ 11}_{-  9}$& $  258^{+ 12}_{-  9}$& $  274^{+ 12}_{-  8}$& $ 11.0^{+1.0}_{-0.8}$& $ 6.40^{+0.38}_{-0.35}$& $0.26^{+0.04}_{-0.02}$& $   8^{+   0}_{-   1}$& $ 164^{+  14}_{-  11}$& $   1969^{+ 126}_{- 107}$& $   -111517^{+   4059}_{-   3464}$\\
Arp 2     &  21.3  & $  242^{+  5}_{-  8}$& $   62^{+ 20}_{- 12}$& $  307^{+  7}_{- 10}$& $ 62.7^{+7.8}_{-7.1}$& $17.69^{+1.00}_{-0.99}$& $0.56^{+0.03}_{-0.03}$& $  79^{+   2}_{-   3}$& $1050^{+ 146}_{- 128}$& $   1158^{+ 342}_{- 199}$& $    -44573^{+   2951}_{-   3125}$\\
NGC 6809   &   4.1  & $ -199^{+  2}_{-  3}$& $   77^{+  7}_{- 13}$& $  220^{+  2}_{-  2}$& $  5.8^{+0.2}_{-0.3}$& $ 1.18^{+0.10}_{-0.14}$& $0.66^{+0.03}_{-0.02}$& $  67^{+   4}_{-   2}$& $  76^{+   3}_{-   4}$& $    270^{+  20}_{-  52}$& $   -154177^{+   1894}_{-   2894}$\\
Terzan 8   &  20.3  & $  273^{+  9}_{-  7}$& $   54^{+ 12}_{- 14}$& $  328^{+  7}_{-  7}$& $ 76.9^{+11.3}_{-11.2}$& $17.48^{+0.63}_{-1.22}$& $0.63^{+0.04}_{-0.03}$& $  82^{+   2}_{-   2}$& $1294^{+ 213}_{- 210}$& $    901^{+ 194}_{- 246}$& $    -39796^{+   3125}_{-   3943}$\\
Pal 11    &   8.7  & $  -20^{+ 13}_{- 18}$& $  152^{+ 13}_{-  7}$& $  153^{+ 14}_{-  7}$& $  8.7^{+0.4}_{-0.3}$& $ 4.20^{+0.56}_{-0.33}$& $0.35^{+0.03}_{-0.05}$& $  27^{+   2}_{-   2}$& $ 122^{+   8}_{-   6}$& $   1183^{+ 121}_{-  82}$& $   -126706^{+   3222}_{-   2403}$\\
NGC 6838   &   7.0  & $   38^{+  5}_{-  6}$& $  204^{+  1}_{-  2}$& $  211^{+  1}_{-  2}$& $  7.3^{+0.2}_{-0.1}$& $ 5.00^{+0.11}_{-0.10}$& $0.18^{+0.02}_{-0.01}$& $  12^{+   1}_{-   0}$& $ 110^{+   4}_{-   1}$& $   1422^{+  28}_{-  24}$& $   -132368^{+   1198}_{-    896}$\\
NGC 6864   &  14.2  & $  -96^{+ 11}_{-  5}$& $   15^{+ 11}_{- 16}$& $  111^{+  6}_{- 10}$& $ 16.0^{+0.8}_{-0.8}$& $ 0.41^{+0.31}_{-0.24}$& $0.95^{+0.03}_{-0.04}$& $  62^{+  40}_{-  17}$& $ 180^{+   7}_{-  11}$& $    169^{+ 114}_{- 177}$& $   -104981^{+   1664}_{-   3080}$\\
NGC 6934   &  12.8  & $ -291^{+ 15}_{-  7}$& $  109^{+ 18}_{- 13}$& $  334^{+  7}_{- 10}$& $ 42.7^{+3.0}_{-3.6}$& $ 2.69^{+0.58}_{-0.38}$& $0.88^{+0.02}_{-0.02}$& $  23^{+   1}_{-   1}$& $ 548^{+  46}_{-  53}$& $   1280^{+ 229}_{- 164}$& $    -61660^{+   2704}_{-   3494}$\\
NGC 6981   &  12.5  & $ -154^{+  7}_{- 14}$& $   -6^{+ 11}_{-  8}$& $  231^{+ 10}_{-  4}$& $ 21.8^{+1.7}_{-0.6}$& $ 0.42^{+0.00}_{-0.27}$& $0.96^{+0.03}_{-0.00}$& $ 114^{+   9}_{-  46}$& $ 244^{+  25}_{-   8}$& $    -50^{+ 100}_{-  73}$& $    -90804^{+   3817}_{-   1536}$\\
NGC 7006   &  36.6  & $ -142^{+  6}_{-  4}$& $  -27^{+ 11}_{-  3}$& $  168^{+  4}_{-  5}$& $ 53.2^{+4.0}_{-1.8}$& $ 2.21^{+0.63}_{-0.85}$& $0.92^{+0.03}_{-0.02}$& $ 130^{+   3}_{-  13}$& $ 708^{+  66}_{-  31}$& $   -909^{+ 374}_{- 141}$& $    -54035^{+   2405}_{-   1196}$\\
NGC 7078   &  10.8  & $    5^{+  8}_{-  8}$& $  121^{+  7}_{-  7}$& $  125^{+  7}_{-  6}$& $ 10.9^{+0.4}_{-0.5}$& $ 3.77^{+0.35}_{-0.35}$& $0.48^{+0.04}_{-0.03}$& $  29^{+   1}_{-   2}$& $ 144^{+   7}_{-   7}$& $   1166^{+  90}_{-  88}$& $   -118429^{+   2182}_{-   2496}$\\
NGC 7089   &  10.6  & $  167^{+  6}_{-  7}$& $  -20^{+  7}_{-  8}$& $  241^{+  7}_{-  9}$& $ 18.9^{+1.1}_{-0.7}$& $ 0.59^{+0.00}_{-0.29}$& $0.94^{+0.03}_{-0.00}$& $ 119^{+  11}_{-  10}$& $ 214^{+  14}_{-   9}$& $   -158^{+  58}_{-  73}$& $    -97367^{+   2754}_{-   1865}$\\
NGC 7099   &   7.4  & $  -48^{+ 14}_{-  9}$& $  -60^{+ 17}_{-  3}$& $  132^{+  4}_{- 11}$& $  8.5^{+0.3}_{-0.2}$& $ 1.00^{+0.06}_{-0.31}$& $0.79^{+0.06}_{-0.01}$& $ 122^{+   2}_{-   6}$& $  98^{+   3}_{-   5}$& $   -245^{+  71}_{-  15}$& $   -135445^{+   1756}_{-   1998}$\\
Pal 12    &  15.2  & $  137^{+ 23}_{- 24}$& $  298^{+ 11}_{- 18}$& $  346^{+ 10}_{- 13}$& $ 58.9^{+11.1}_{-9.2}$& $15.01^{+0.64}_{-0.43}$& $0.59^{+0.05}_{-0.05}$& $  67^{+   1}_{-   1}$& $ 948^{+ 193}_{- 150}$& $   2020^{+ 107}_{- 119}$& $    -46965^{+   4540}_{-   4639}$\\
Pal 13    &  24.6  & $  266^{+  6}_{-  7}$& $  -62^{+  8}_{- 11}$& $  284^{+  6}_{-  6}$& $ 71.2^{+4.1}_{-4.5}$& $ 6.53^{+0.61}_{-0.44}$& $0.83^{+0.01}_{-0.02}$& $ 112^{+   5}_{-   4}$& $1044^{+  73}_{-  77}$& $  -1156^{+ 197}_{- 250}$& $    -44251^{+   1569}_{-   1857}$\\
NGC 7492   &  23.6  & $  -70^{+ 14}_{- 12}$& $  -11^{+  3}_{-  5}$& $  101^{+ 11}_{-  8}$& $ 26.1^{+1.4}_{-0.9}$& $ 1.73^{+0.59}_{-0.63}$& $0.88^{+0.04}_{-0.04}$& $  96^{+   3}_{-   1}$& $ 314^{+  20}_{-  14}$& $    -97^{+  27}_{-  43}$& $    -81114^{+   2319}_{-   1619}$\\
\hline
 \end{tabular}
   \end{center}
   \end{scriptsize}
  \end{minipage}}

  \begin{figure*}
{\begin{center}
   \includegraphics[width=1.4\textwidth,angle=0]{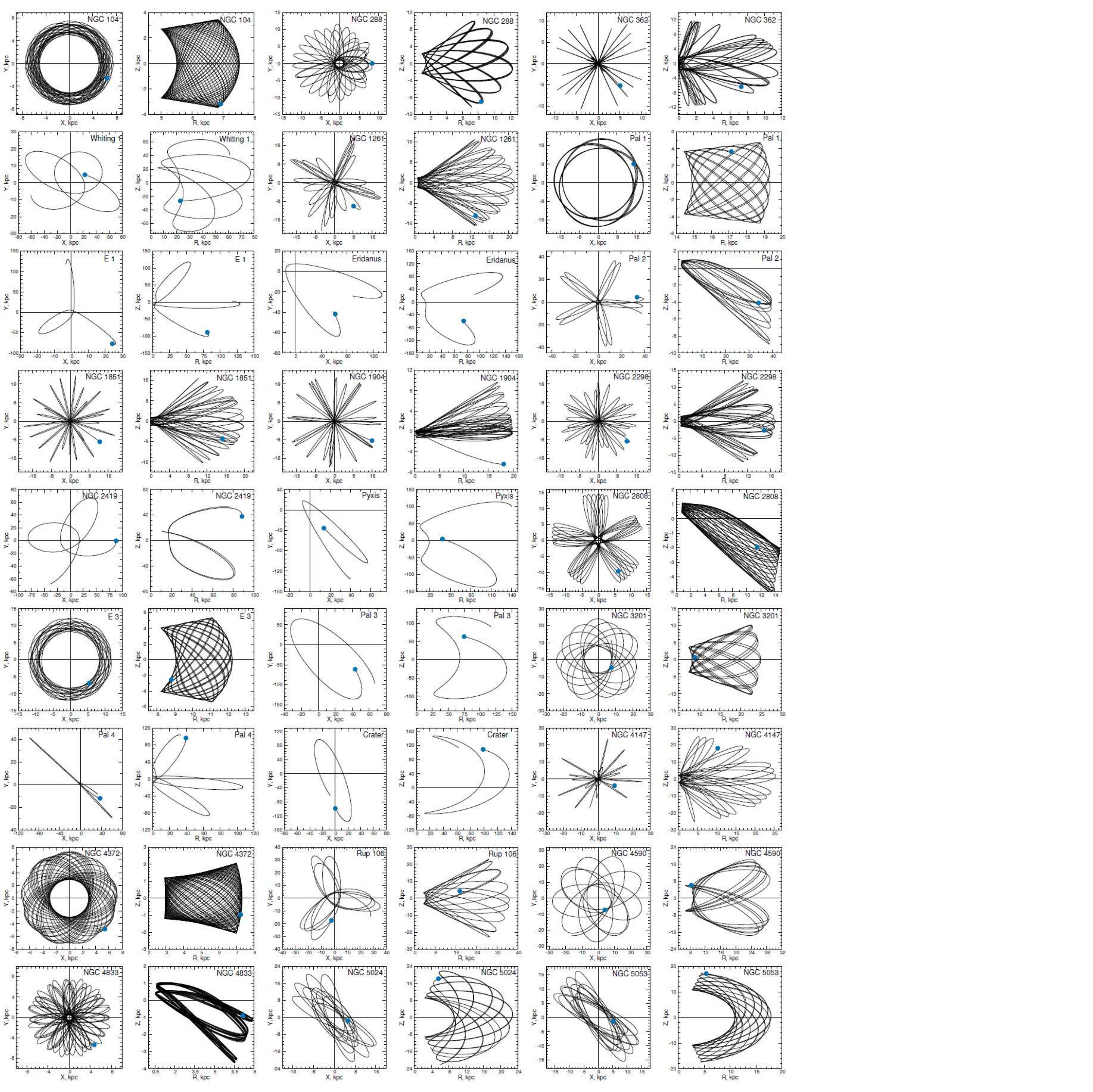}
    \caption{Orbits of the GC of the Milky Way galaxy in two projections $(X,Y)$ and $(R,Z)$. The blue circle marks the beginning of the orbit.}
\label{fcat}
\end{center}}
\end{figure*}

  \begin{figure*}
{\begin{center}
   \includegraphics[width=1.4\textwidth,angle=0]{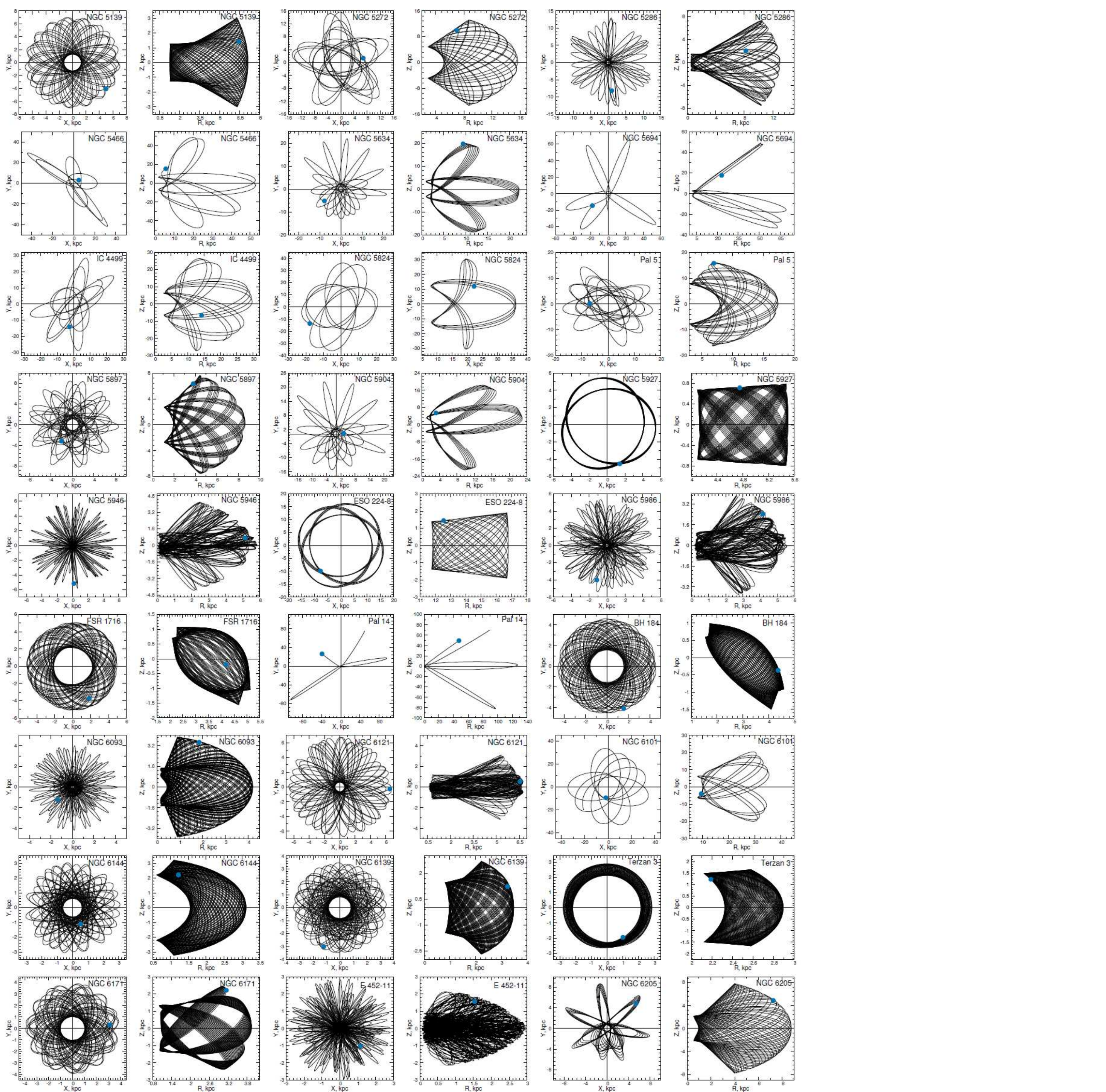}
\centerline{Figure 5: Continued}
\label{fcat}
\end{center}}
\end{figure*}

  \begin{figure*}
{\begin{center}
   \includegraphics[width=1.4\textwidth,angle=0]{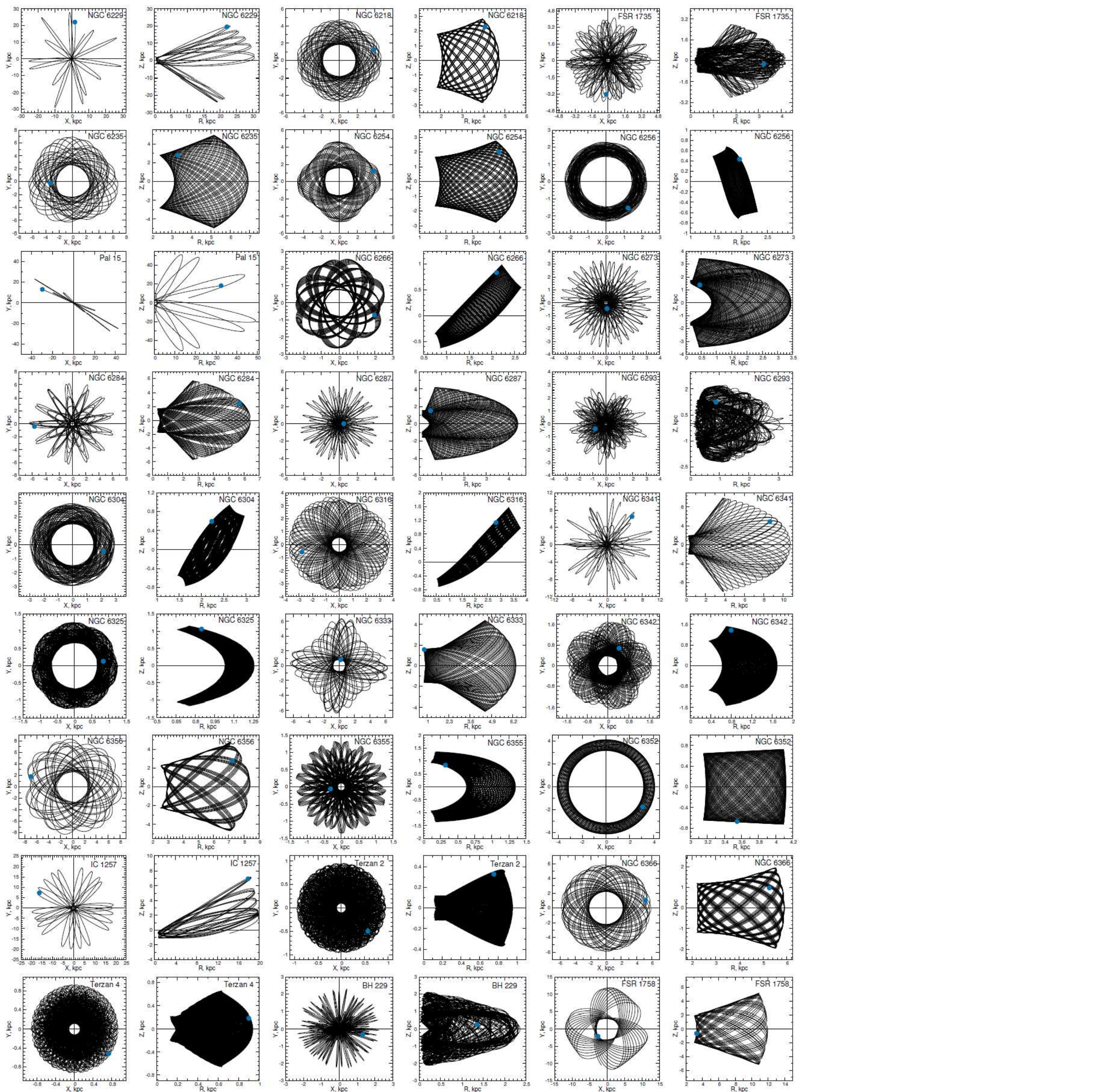}
\centerline{Figure 5: Continued}
\label{fcat}
\end{center}}
\end{figure*}

  \begin{figure*}
{\begin{center}
   \includegraphics[width=1.4\textwidth,angle=0]{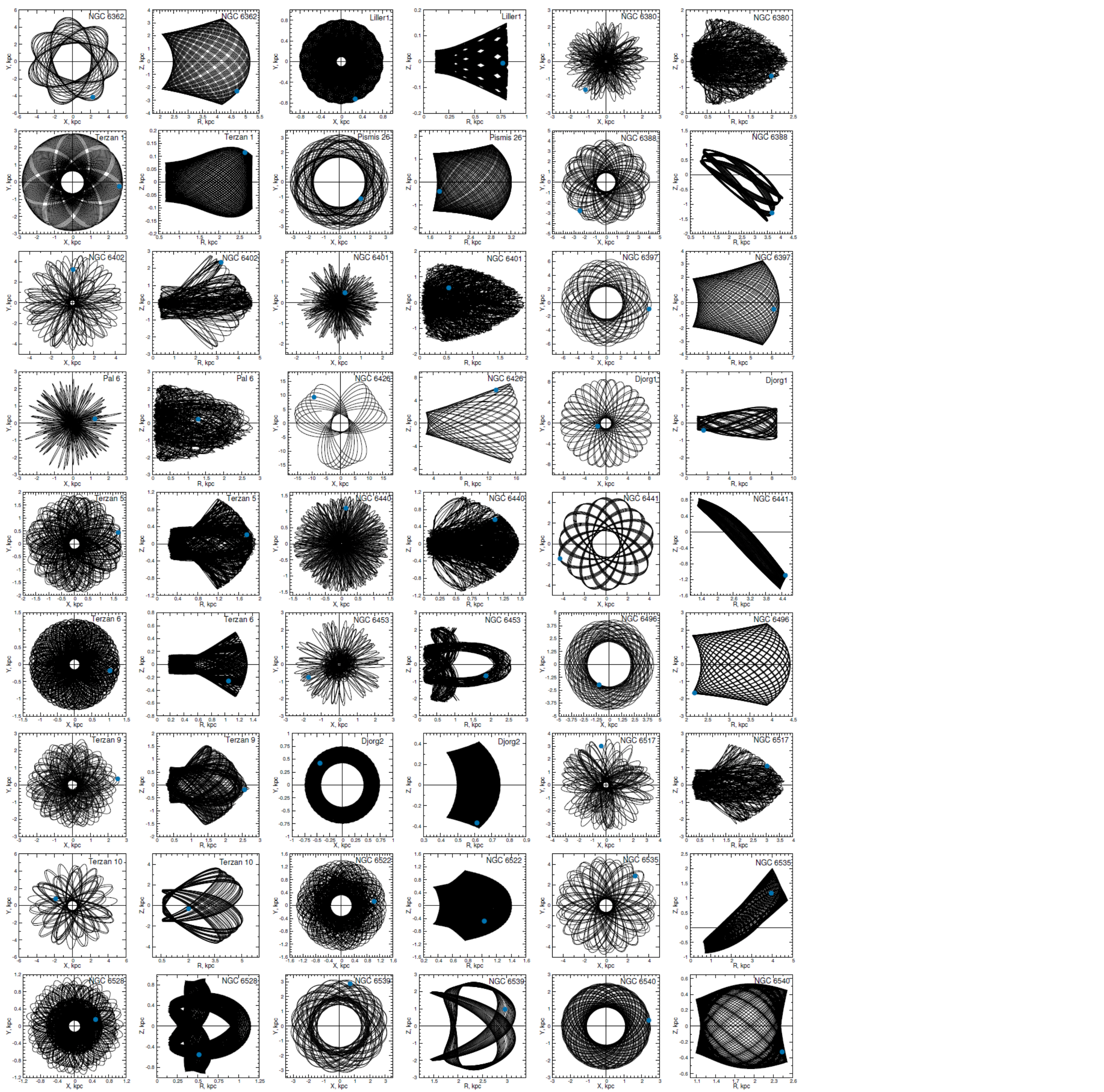}
\centerline{Figure 5: Continued}
\label{fcat}
\end{center}}
\end{figure*}

  \begin{figure*}
{\begin{center}
   \includegraphics[width=1.4\textwidth,angle=0]{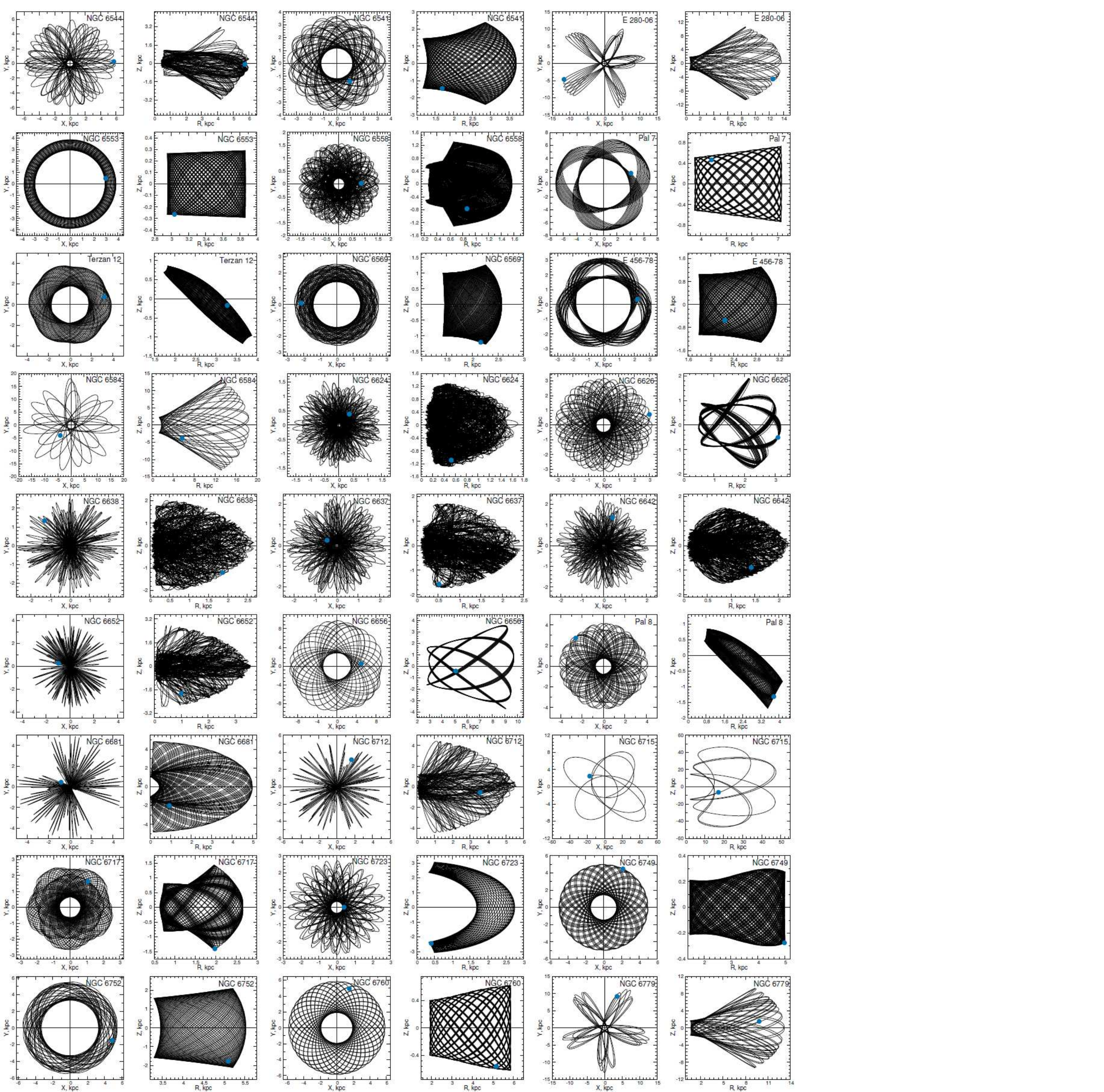}
\centerline{Figure 5: Continued}
\label{fcat}
\end{center}}
\end{figure*}

  \begin{figure*}
{\begin{center}
   \includegraphics[width=1.0\textwidth,angle=0]{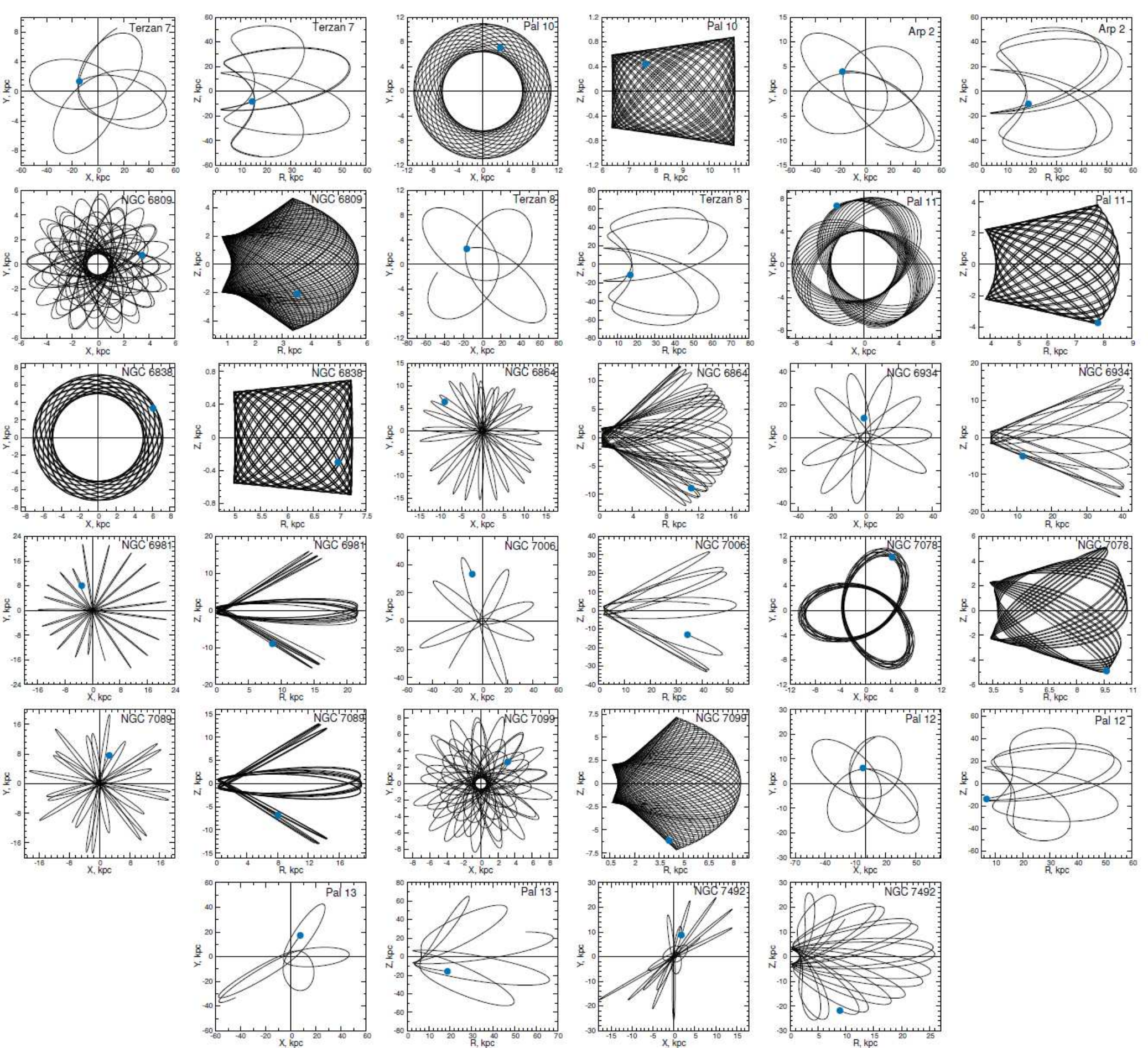}
\centerline{Figure 5: Continued}
\label{fcat}
\end{center}}
\end{figure*}

\end{document}